\documentclass[preprints,article,accept,pdftex,moreauthors]{Definitions/mdpi}
\usepackage{journals}

\firstpage{1} 
\pubvolume{1}
\issuenum{1}
\articlenumber{0}
\pubyear{2025}
\copyrightyear{2025}
\externaleditor{Lorenzo Iorio} 
\datereceived{7 April 2026} 
\daterevised{11 June 2026} 
\dateaccepted{30 June 2026} 
\datepublished{ } 

\Title{Projection-Enhanced Disk Breaks: Evidence from Deep Photometric Decomposition}

\Author{Sergey S. Savchenko
$^{1,2,}$*\orcidA{},
  Ilia V. Chugunov $^{1,3}$\orcidB{},
  Alexander A. Marchuk $^{1,2}$\orcidC{},
  Vladimir P. Reshetnikov $^{1,2}$\orcidD{},
  \mbox{Matvey D. Kozlov $^{1,2}$\orcidH{}},
  Dmitry I. Makarov $^{1,4}$\orcidE{},
  Aleksandra V. Antipova $^{1,4}$\orcidF{},
  Anastasia M. Sypkova $^{1,2}$\orcidG{}
}

\AuthorNames{Sergey S. Savchenko, Ilia V. Chugunov, Alexander A. Marchuk , Vladimir P.
Reshetnikov, Matvey D. Kozlov, Dmitry I. Makarov, Aleksandra V. Antipova
and Anastasia M. Sypkova}

\address{%
$^{1}$ \quad Central (Pulkovo) Astronomical Observatory of RAS, Pulkovskoye Chaussee 65/1, \mbox{196140 St. Petersburg, Russia};\\
$^{2}$ \quad Saint Petersburg State University, 7/9 Universitetskaya nab., 199034 St.Petersburg , Russia \\
$^{3}$ \quad Sternberg Astronomical Institute, Lomonosov Moscow State University, Universitetsky Pr. 13, \mbox{119234 Moscow, Russia}\\
$^{4}$ \quad Special Astrophysical Observatory, Russian Academy of Sciences, 369167 Nizhnii Arkhyz, Russia \\
}

\corres{Correspondence: s.s.savchenko@spbu.ru}

\abstract{Radial brightness profiles of disk galaxies often exhibit so-called breaks---locations where their
  exponential-scale length abruptly changes. Some galaxies have downbending (Type~II) breaks, where their brightness
  decays faster in outer regions, while other have upbending (Type~III) breaks, resulting in more extended outer disks
  or envelopes. Disk radial profiles without any breaks (Type~I) appear to constitute a minority. The exact fractions of
  different break types depend on many galactic parameters---such as Hubble type, stellar mass, spatial environment, and
  bar presence---and vary significantly across different studies. Another source of discrepancy is the orientation of
  galaxies: projection effects may play an important role in break detectability. In this work, we utilize DESI Legacy
  DR10 imaging to perform photometric decomposition of a sample of 375 edge-on galaxies and investigate their radial
  breaks. We find that the vast majority ($\approx$90\%) of disks in our sample have Type~II breaks, which is a
  considerably higher fraction than in many previous works ($\sim$50\%). We carefully tested our results to check if
  observed breaks can be a result of flaring or two-disk composition. We showed that a high fraction of Type~II breaks
  can be attributed to projection effects, which enhance the observed surface brightness of breaks in edge-on galaxies.}

\begin{document}
\section{Introduction}
In the local Universe, about three quarters of bright galaxies have a disk morphology~\citep{Conselice2006, Kelvin2014}.
Their general structure can be reasonably well described by two distinct stellar components: a central concentration of
stars (bulge) and a rotationally supported flat stellar disk (the latter often hosts other sub-components such as bars
and spiral arms). To~measure the properties of these components, the~method of photometric decomposition is often used
(see, for~example, \citep{Peng2010, Erwin2015}). The~goal of this method is to describe galactic components through
analytical functions and to analyze the best-fitting parameters in order to infer the components'
properties. To~describe the spatial light distribution of a bulge, the~usual choice is a single S{\'e}rsic function
\citep{Sersic1963}. Disks, on~the other hand, often exhibit more complex behavior. In~the pioneering works
\citep{Patterson1940, deVac1959}, the~exponential law was shown to be a good approximation to the radial brightness
distribution in a disk component. Even though a simple exponential law is still often used in modern studies on this
topic as a default option to describe disks, it was shown in \citep{Freeman1970, vanderKruit1979, Erwin2008,
  Silchenko2009} and many later works that a considerable fraction of disks cannot be fitted properly with this
simple~function.

The most common feature in the disk radial brightness distribution is the so-called break: a region where the radial
exponential-scale length changes abruptly.  If~the profile becomes steeper beyond the break radius (i.e., if the
exponential-scale length decreases), the~galaxy is said to have a Type II profile, also known as a downbending or
truncated profile.  In~the opposite case, the~galaxy has a Type~III profile (upbending or antitruncated). If~no breaks
appear, the~galaxy has a Type~I profile (pure exponential).

Substantial effort has been made to explain the mechanisms of break formation. If~a galaxy forms from the collapse of a
uniform rotating gas sphere, the~resulting disk will have a roughly exponential radial density distribution with a
truncation located at about \mbox{4.5 exponential} scales from its center \citep{Mestel1963}. Despite its idealized
assumptions, the~model demonstrates that Type~II radial disk breaks can emerge naturally during secular disk
evolution. A~downside of this model is that the present-day angular momentum distribution in galactic disks may have
little in common with that of a protogalactic disk. Non-axisymmetric features such as spiral arms \citep{Sellwood2002,
  Roskar2008} and bars \citep{Sellwood1980, Athanassoula2003, Debattista2006} can effectively redistribute angular
momentum in disks, especially at radii close to corotation or Lindblad resonances. This redistribution should smear out
various radial features in disks, such as radial gradients of ages or metallicities, including peculiarities of the
radial brightness profile. On~the other hand, stars near the outer Lindblad resonance (OLR) of a bar effectively
exchange angular momentum with it and migrate to larger, nearly round orbits forming rings with radii 2-3 times larger
than the bar size~\cite{MunozMateos2013}. Those rings emerge as a bumpy feature in the azimuthally averaged radial
profiles of disks and may appear as a Type~II break on a profile (such breaks are dubbed Type II-OLR in
\citep{Erwin2008}).

Another explanation for Type~II break formation is that they may be related to an abrupt drop in a star-formation rate
beyond the break radius. Such breaks are sometimes called a truncation. Both theoretical \citep{Goldreich1965} and
empirical \citep{Kennicutt1989} considerations lead to the result that beyond some critical radius, the~star-formation
rate should decrease rapidly when the gas stability criterion $Q_g(R)$ exceeds unity or the gas surface density
$\Sigma_g(R)$ falls below the critical value. This model is supported by both simulations and observations of real
galaxies. For~example, Ref. \citep{Roskar2008} carried out a simulation of a star-forming disk evolution. In~this
simulation, a~Type~II break appears almost immediately due to a sharp drop in star-formation at the radius, where the
gas has not yet cooled enough to effectively form stars. Most of the stars beyond the break were formed in the inner
regions and were stirred outward beyond the break by interaction with spiral arms. This process should lead to a
U-shaped profile of stellar ages and colors in disks, with~a minimum located near the break radius. Indeed, such color
profiles are observed in many galaxies \citep{Bakos2008, Camba2022, Raji2025}. It was also noted that the occurrence of
Type~II breaks is related to the star-formation rate: break frequency increases toward later Hubble types
\citep{Pohlen2006, Erwin2008, Gutierrez2011, Laine2016, MendezAbreu2017, Tang2020}, whereas galaxies with suppressed
star-formation (for example, due to the ram pressure stripping or other effects in dense environments) tend to have a
reduced fraction of breaks \citep{Erwin2012, Head2015, Pranger2017, Silchenko2018, Silchenko2020, Pfeffer2022,
  Mondelin2025}. Modern cosmological simulations allow one to track the evolution of radial profiles of stellar and gas
surface density along with the star-formation rate on cosmological timescales. In~\citep{Chen2026} a sample of model
galaxies from the Illustris TNG-50 simulation \citep{Nelson2019} was investigated to determine how Type~II breaks form
in them. It was shown that the peak of the specific star-formation rate appears near the present-day break radius at
$z\approx 0.5$, which leads to an enhanced star assembly compared to both inner and outer regions, effectively producing
a break. The~same process also reproduces the U-shaped radial curve of stellar ages with the local minimum near the
break~radius.

There appears to be even less agreement on the explanation of Type~III (upbending) breaks. In~\citep{Erwin2005,
  Pohlen2006} it was noted that Type~III breaks can be divided into two subclasses: Type~III-s and
Type~III-d. The~distinctive feature of the first subclass is that isophotes become rounder beyond the break radius,
which could mean that the excess of light beyond the break is not related to the disk itself, but~rather to some
spherical component such as a stellar halo \citep{Bakos2012} or even the extended wings of a bright bulge
\citep{Maltby2012}. In~the case of Type~III-d, the~isophotes beyond the break radius have the same shape, which may
indicate that this feature is associated with the disk. The~behavior of the outer isophotes, on~the other hand, is not
an unambiguous classification parameter, since for face-on galaxies the isophotes of both disk and spherical components
would have the same~shape.

The observed frequency of Type~II breaks varies significantly among studies: $12^{+3}_{-2}\%$~\citep{Head2015}, $25\%$
\citep{Erwin2012}, $42\%$ \citep{Tang2020}, $48\%$ \citep{Erwin2008}, $50\pm 4\%$ \citep{Gutierrez2011}, $66\%$
\citep{Pohlen2006}, $82\pm 16\%$ \citep{MartinNavarro2012}. This discrepancy can be caused by various factors: different
photometric depths of used imaging, different photometric bands used, differences in sample distribution by Hubble type,
spatial environment, stellar mass, bar presence, etc. Another reason is the orientation of galaxies: some samples mainly
consist of galaxies oriented close to a face-on orientation, while others contain only edge-on galaxies. Galaxy
orientation may greatly impact the detectability of breaks due to projection effects and different line-of-sight
integration paths leading to different surface brightness values around break~points.

In this work, we investigate a sample of 375 edge-on galaxies in order to determine the occurrence rate of Type~II
breaks, and we compare these results with previous measurements made for face-on or mildly inclined galaxies by other
authors.  This will allow us to make a bridge between two distinct samples of galaxies: face-ons and edge-ons.  As a
source of images we use the DESI Legacy DR10 survey ($g$, $r$, and~$i$ bands) \citep{Dey2019}, which is about
1--2~magnitudes deeper than the SDSS~\cite{Abdurrouf2022} survey often used in previous works on this topic.  This
article is organized as follows. In~Section~\ref{sec:breaks_fon_eon}, we discuss the implications of projection effects
for the observation of breaks in disks viewed at different angles, including face-on and edge-on orientations.  In
Section~\ref{sec:sample}, we describe our sample of galaxies. In~Section~\ref{sec:method}, we briefly describe our
approach to the decomposition. Our results are presented in Section~\ref{sec:results}, while our main conclusions are
summarized in Section~\ref{sec:conclusions}.

\section{Breaks in Face-on and Edge-on Disks}
\label{sec:breaks_fon_eon}
In this section we consider how breaks appear when projected onto the sky plane when a disk is viewed in different
orientations, especially in edge-on and face-on orientations as special cases. A~simple transparent Type~I disk with an
exponential profile in both the radial and vertical directions has the following luminosity density as a function of
cylindrical coordinates $R$ (distance to the disk center) and $z$ (height above or below the disk plane):
\begin{equation}
  \rho(R, z) = \rho_0 \exp \left( -\frac{R}{h_r} \right) \exp\left( -\frac{\left| z \right|}{h_z} \right).
  \label{eq:simple_disk}
\end{equation}

When viewed face-on, the~line-of-sight integration is along the $z$-axis and reduces the second exponential, so that the
observed surface brightness distribution projected onto the sky plane becomes
$$
\Sigma_{\mathrm{face}}(R) = 2h_z\rho_0 \exp \left ( -\frac{R}{h_r} \right).
$$

In the edge-on orientation, the~line-of-sight integration is parallel to the disk plane and leads to an Abel transform of
a radial part of the density distribution:
$$
\Sigma_{\mathrm{edge}}(R) = 2 \int_R^\infty \rho_0 \exp \left ( -\frac{x}{h_r} \right) \frac{x}{\sqrt{x^2 - R^2}}dx=2\rho_0 R K_1\left (\frac{R}{h_r} \right),
$$
where $K_1$ is a Bessel function of the second kind \citep{vanderKruit1981}.

If there is a sharp break in the radial law of the volume brightness density, the~equation for $\rho (R, z)$ becomes:
\begin{equation}
  \rho(R, z) = \left\{
    \begin{array}{l}
      {\rho_{0, 1}\cdot \exp\left(-\frac{R}{h_1} \right)\exp\left( -\frac{\left| z \right|}{h_z} \right),\quad R < R_b}\\
      {\rho_{0, 2}\cdot \exp \left(-\frac{R}{h_2}\right)
      \exp\left( -\frac{\left| z \right|}{h_z} \right),\quad R > R_b,}
    \end{array}
  \right.
  \label{eq:broken_disk}
\end{equation}
where $\rho_{0, 1}$ is the central volume density of an inner disk,
and~$\rho_{0, 2}=\exp\left (R_b\left ( \frac{1}{h_2} - \frac{1}{h_1} \right)\right)$ is a central volume density of an
outer disk, computed under the assumption that there is no discontinuity at the break radius. The shape of a broken
  disk is governed, therefore, by~two radial scales, an~inner one $h_1$ and an outer one $h_2$. The~break
  strength can be expressed either by a linear coefficient between inner and outer radial scales ($h_2 = \alpha h_1$)
or~on a logarithmic scale by defining strength as
\begin{equation}
  s = \log_{10}\left(\frac{h_2}{h_1}\right)
  \label{eq:strength}
\end{equation}
(see, e.g.,~\citep{Maltby2012, Borlaff2016}). Negative values correspond to Type~II breaks and positive values to Type~III breaks, and~the larger the absolute value, the~more pronounced the break~is.

Again, when viewed face-on, the~line-of-sight integration along the vertical direction of the disk yields a projected surface brightness distribution
with an exponential decay in the radial direction and a sharp break:
\begin{equation}
  \Sigma_{\mathrm{face}}(R) = \left\{
    \begin{array}{l}
      {2h_z\rho_{0, 1}\exp\left(-\frac{R}{h_1}\right), \quad R<R_b}\\
      {2h_z\rho_{0, 2}\exp\left(-\frac{R}{h_2}\right), \quad R>R_b.}
    \end{array}
  \right.
  \label{eq:fon_integration_with_breaks}
\end{equation}

\begin{figure}[H]
  \centering
  \includegraphics[width=0.4\linewidth]{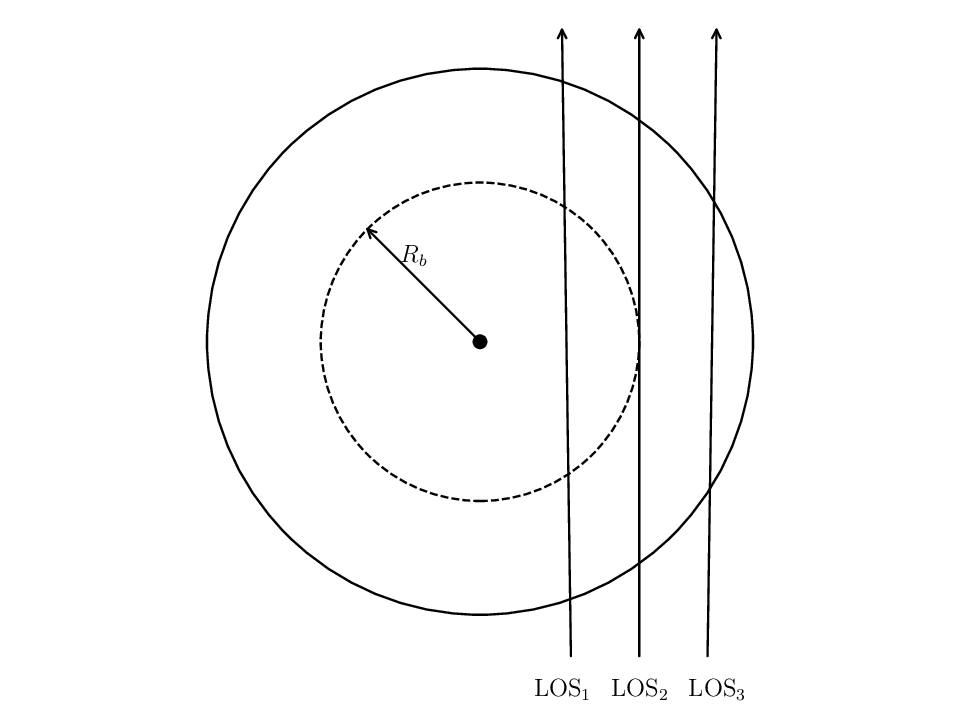}
  \caption{A schematic face-on view of a galaxy, with~three parallel lines of sight piercing it at three distinct
    points: $\mathrm{LOS_1}$ lies inside the break radius, $\mathrm{LOS_2}$ passes exactly through the break radius,
    and~ $\mathrm{LOS_3}$ lies outside of the break radius. It is easy to see that the surface brightness at the
    locations of $\mathrm{LOS_2}$ and $\mathrm{LOS_3}$ is governed only by the outer disk density distribution.}
  \label{fig:los}
\end{figure}

The picture becomes more complicated when a disk is viewed edge-on. In~this case, the~line-of-sight integration along
the disk midplane inside the break radius mixes light from the inner and outer disks. Figure~\ref{fig:los} schematically
illustrates this effect. This figure shows a view of a disk that has a break at the break radius $R_b$ and three lines
of sight: the one that lies inside the break radius ($\mathrm{LOS_1}$), the~one that passes exactly through the break
radius ($\mathrm{LOS_2}$), and~the one that lies outside the break radius ($\mathrm{LOS_3}$). Thus, $\mathrm{LOS_1}$
integrates fluxes from both the inner and outer disks, whereas $\mathrm{LOS_2}$ and all outer lines of sight, including
$\mathrm{LOS_3}$, integrate only through the outer~disk.

Analytically this can be expressed in the following form:

\vspace{-9pt}

\begin{adjustwidth}{-\extralength}{0cm}
\centering 
\begin{equation}
  \Sigma_{\mathrm{edge}}(R) = \left\{
    \begin{array}{l}
      {2 \int_R^{R_b}\rho_{0,1}\exp\left(-\frac{x}{h1}\right)\frac{x}{\sqrt{x^2-R^2}}\,dx +
      2 \int_{R_b}^\infty\rho_{0,2}\exp\left(-\frac{x}{h2}\right)\frac{x}{\sqrt{x^2-R^2}}\,dx, \quad R<R_b} \\
      {2 \int_{R}^\infty\rho_{0,2}\exp\left(-\frac{x}{h2}\right)\frac{x}{\sqrt{x^2-R^2}}\,dx, \quad R>R_b.}
    \end{array}
  \right.
  \label{eq:eon_integration_with_breaks}
\end{equation}
\end{adjustwidth}

Two integrals (for the inner and outer disks) contribute in the region $R<R_b$, while only the outer disk contributes in
the region $R>R_b$. Unfortunately, the~integrals in (\ref{eq:eon_integration_with_breaks}) cannot be solved analytically
and for $R < R_b$ they cannot be reduced to well-documented Bessel functions. Nevertheless, some important conclusions
can be drawn regarding the appearance of breaks in face-on vs. edge-on~orientations.

\begin{figure}[H]
  \centering
  \includegraphics[width=0.6\linewidth]{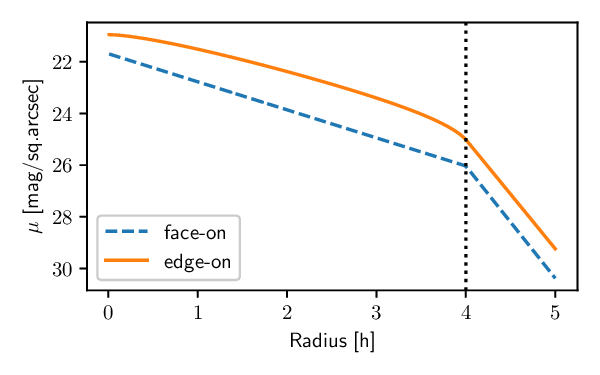}
  \caption{Radial surface brightness for an edge-on and a face-on disk with the same parameters (surface brightness
    of a model scaled such that for a face-on disk, $\mu (0)=21.7$ mag/sq.arcsec).}
  \label{fig:eon_fon_surfbri_compar}
\end{figure}

First, in~the edge-on orientation the break appears smoother and its observed location shifts towards the galactic
center. This happens because inside the true break radius ($\mathrm{LOS_1}$ in Figure~\ref{fig:los}), a~weighted average
of the inner and outer disks is observed, so the effective observed exponential scale lies somewhere between the scale
lengths of the inner and outer disks. As~$R$ increases, the~weight of the inner disk decreases, and~the outer disk
starts to dominate the total observed flux. At~$R=R_b$ and farther from the galactic center ($\mathrm{LOS_2}$ and
$\mathrm{LOS_3}$ in Figure~\ref{fig:los}), only the outer disks contributes to the observed surface brightness. This means
that the observed transition from the inner to the outer disk in the radial surface brightness distribution should start
before $R_b$ and finish at $R_b$. To~illustrate this effect, the~radial surface brightness distributions for the
same disk with $h_2 = 0.25 h_1$ and $R_b=4h_1$ are plotted in Figure~\ref{fig:eon_fon_surfbri_compar} for the face-on
and edge-on orientations. It is seen that the transition between the inner and outer disks in the edge-on orientation
appears smoother and occurs somewhat inside the true value of $R_b$. This implies that simple linear piecewise fits to
two disks should lead to underestimation of the $R_b$ value (along with the overestimation of radial-scale lengths,
as noted in \citep{deGrijs1998, Kregel2002}).

The second point is that, due to projection effects, disks in edge-on orientation should have higher values of surface
brightness at the same galactocentric distances than the same disks viewed face-on (at least if one considers the disks
to be transparent). The~difference is larger for thinner galaxies because for them the face-on line-of-sight integration
results in lower values. By~plugging $R_b$ into (\ref{eq:fon_integration_with_breaks}) and
(\ref{eq:eon_integration_with_breaks}) and taking the ratio, one can find how much the surface brightness at the break
point is higher in the edge-on orientation than when the same disk is viewed face-on:
\begin{equation}
\frac{\Sigma_{\mathrm{edge}}(R_b)}{\Sigma_{\mathrm{face}}(R_b)} = 
\frac{R_b}{h_z}K_1\left(\frac{R_b}{h_2}\right)\exp{\left(\frac{R_b}{h_2}\right)}.
\label{eq:sb_diff_in_break}
\end{equation}

For example, for~a disk with parameters $h_2=0.25h_1$, $R_b=4h_1$, and~$h_z=0.2h_1$, which are common for disk galaxies,
this equation gives the ratio $\Sigma_{\mathrm{edge}}(R_b)/\Sigma_{\mathrm{face}}(R_b)=6.4$, which corresponds to a
difference of about two magnitudes. This amplification of the surface brightness of the break point in the edge-on
orientation relative to the face-on orientation depends on the shape of the outer disk. All else being equal, a~more
extended outer disk gives higher observed surface brightness at the break point when viewed edge-on (see
  Appendix~\ref{ap:break_brightness} for more details). This should lead to a systematic shift: in edge-on disks,
stronger downbending breaks (lower $h_2$ values) should appear, on~average, at~lower surface brightnesses, whereas
stronger upbending breaks (higher $h_2$ values) should appear at higher surface brightnesses.

\begin{figure}[H]
  \centering
  \includegraphics[width=0.6\linewidth]{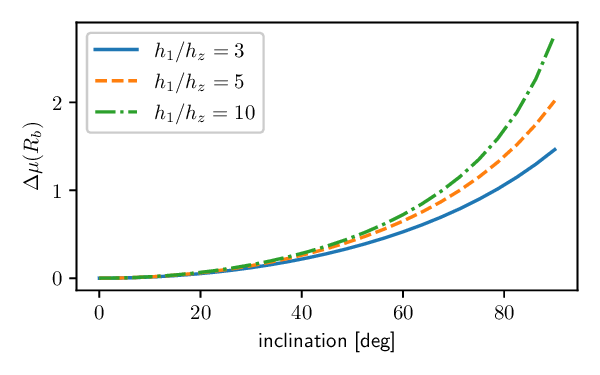}
  \caption{The difference in surface brightness at the break point between an inclined disk and the same disk in the
    face-on orientation. Different lines correspond disks with different intrinsic thicknesses (see legend).}
  \label{fig:r_break_brightness_on_inclination}
\end{figure}

This result cannot be generalized analytically for arbitrary inclinations, but~a numerical estimation is possible.
Figure~\ref{fig:r_break_brightness_on_inclination} shows how the observed surface brightness at the break point depends on
inclination for disks with three different ratios of radial to vertical scale (the remaining parameters are the same as
in the example above). As~expected from Equation~(\ref{eq:sb_diff_in_break}), thinner disks exhibit a larger
ratio of edge-on to face-on surface brightness at the break point (the rightmost points of the curves), and~it is clear from
the plot that this is true for all inclinations. It is also important to note that the curves are non-linear, and~the greatest increase occurs near the high-inclination~end. 

\vspace{-2pt}

In the conclusion of this section we would like to note that the approach described above, along with the decomposition
method (Section~\ref{sec:method}), treats a galaxy only within the framework of the observed brightness
distribution. But~the underlying physical quantity is the volume mass density of stars (or the surface mass density when
the galaxy is projected on the sky plane). The~direct measurement of the volume mass density from the observed
projection of a light distribution is not an easy task for several reasons. First, an~inverse Abel transform or other
deprojection approach would be necessary to solve this problem \citep{Pohlen2007}. Decomposition using 3D brightness
distribution functions as performed in this work can also give information about the intrinsic volume density
distribution, but~the result becomes model- dependent. In any case, in~the second stage, a~deprojected 3D brightness
distribution should be converted into a mass distribution using a mass-to-light rule (like \citep{Bell2003,
  Ebrova2025}), which becomes an additional source of uncertainty, especially if the conversion is performed using
optical magnitudes. Still, if~performed carefully, this would provide interesting insights about disks, since it has
already been proven that their brightness and mass distributions do not necessarily coincide \citep{Bakos2008,
  Bakos2012, Tang2020}. But~this kind of analysis goes beyond the scope of this article.

\section{The~Sample}
\label{sec:sample}
The main task of this study---the detection of disk breaks in edge-on galaxies---requires a statistically significant
sample of galaxy images. Since breaks are usually observed in the outer regions of disks, where the surface brightness
is relatively low (and beyond the Type~II break a disk fades out even faster), we decided to use the DESI Legacy Survey
\url{https://www.legacysurvey.org/} \citep{Dey2019} as our source of images. By~computing statistics in randomly
selected $10\times 10$ arc-second boxes in the object-free areas of random fields, we estimate the $3\sigma$ limiting
depth in the $g$-band to be $28.98 \pm 0.41$ mag per sq. arcsec (specified uncertainty is a three-sigma standard
deviation of the limiting depth computed over a sample of random fields).

We build our sample using a combination of the EGIPS \citep{Makarov2022} and RFGC \citep{Karachentsev1999} catalogs of
galaxies viewed close to edge-on. The~EGIPS catalog was created using an artificial neural network classifier applied to
the Pan-STARRS1 survey \citep{Chambers2016}. We extended this catalog with the RFGC galaxies to cover the southern part
of the sky, since the Pan-STARRS1 survey only contains regions of the sky north of $\delta=-30^\circ$. To~refine this
combined catalog and obtain a smaller sample of galaxies viewed very close to edge-on orientation, we performed a visual
examination of the images (The EGIPS database contains information about neural network scores that could in principle
be used as a starting point to exclude galaxies that deviate from edge-on orientation, but~the RFGC catalog does not
have such a parameter, so for consistency we decided to opt for visual inspection of both parent samples). We downloaded
RGB images of all EGIPS galaxies from the DESI Legacy footprint, and~each galaxy was inspected by five randomly selected
participants from the author list. The~decision was made based on the symmetry of the bulge appearance (in an edge-on
flat galaxy the bulge should be divided into two equal parts by the disk plane), the~dust lane shape (should appear as a
straight line and divide the stellar disk into equal parts), and~the overall appearance of outer isophotes (slightly
inclined galaxies demonstrate rounder isophotes in their outer regions). If~a galaxy received at least four positive
votes out of five, we considered it a genuine edge-on galaxy and included it in our sample. This selection left us with
375 galaxies (301 from EGIPS and 74 from RFGC). According to the EGIPS database, the~median $r$-band axis ratio of our
sample galaxies that are included in EGIPS is $0.168$, which is somewhat lower than the median axis ratio of the full
EGIPS sample, which is $0.210$. The~lower value of the axis ratio is a result of stricter selection of galaxies
regarding their proximity to edge-on orientation. We have not performed any inclination measurements of our galaxies
since it is not an easy task for galaxies that are almost edge-on: a usual approach that relies on comparing some
average intrinsic axis ratio to an observed one fails when the inclination is close to 90 degrees. We compared the
apparent shifts in dust lanes of our galaxies to those of edge-on galaxies that have inclination estimates from
radiative transfer decomposition (for example, ~\cite{Bianchi2007, DeGeyter2014}) and concluded that for most of our
galaxies the deviation from true edge-on orientation is within $\approx$1 degree.

We downloaded FITS images of all sample galaxies in the $g$-, $r$-, and~$i$-bands along with corresponding inverse
variance maps via the DESI Legacy cutout service \url{https://www.legacysurvey.org/viewer/cutout.fits}. We also
downloaded the appropriate PSF kernels from the survey web server to build PSF images following the survey guidelines
\url{https://www.legacysurvey.org/dr10/psf/}. To~exclude background and foreground objects and image artifacts from
further analysis, we created mask images using object catalogs generated by the {\small SEXTRACTOR} package
\citep{Bertin1996}. For~every passband we ran {\small SEXTRACTOR}, loaded its output object catalog as a set of
ellipses, excluded the central object (the target galaxy) and multiplied the sizes of all ellipses by 1.5 to properly
cover even the outer low-surface-brightness extended wings of bright objects that can be missed by {\small
  SEXTRACTOR}. We displayed the ellipses overlaid on the galaxy image for a visual inspection and possible modifications
using the SAOImage DS9 package \citep{Joye2003}. After~that we made a binary mask that excludes all pixels inside the
ellipses. As~the final mask we used the union of all masks created for all photometric~bands.

As the final step of the data preparation, we created dust masks to exclude regions of disks heavily contaminated by the
dust lane. Dust can greatly affect the observed parameters of galaxies \citep{Gadotti2010, Pastrav2013a, Pastrav2013b},
especially for edge-on galaxies \citep{Savchenko2023}. Excluding dust-attenuated regions from the analysis is a possible
way to reduce the influence of dust (see, for~example,~\citep{Bizyaev2004, Savchenko2023}). In~this work, we decided to
manually make masks using the SAOImage DS9 package to exclude dust-contaminated regions in the~images.

\section{The~Decomposition}
\label{sec:method}
To obtain information about disk breaks in our sample galaxies, we used the algorithm of photometric decomposition. The~goal of this approach is to construct a model of a galaxy image using a set of analytical functions, such that each
function corresponds to a galactic component. The~values of the galactic parameters are inferred from the best-fitting
parameters of the model. To~do this, we used the {\small IMFIT} package \citep{Erwin2015}. This package takes as inputs a galaxy image, a~weight map, a~masked pixel map, a~PSF image, and~a configuration file describing a model and the initial parameters, and~performs a search for best-fitting
values to these~parameters.

For our models we adopted the following functions. To~describe a 
central light concentration, we used a S{\'e}rsic function \citep{Sersic1963}. The nature of this central light concentration is somewhat uncertain since it may include multiple different components such as classical bulge, pseudo-bulge, bar, nuclear disk, X-structure, etc., and~it
is often difficult to separate them based solely on photometry. Instead, they are often fitted as a single combined component, sometimes called a 'photometric bulge' \citep{Gadotti2026}. Probably the most prominent additional component (besides the bulge/pseudo-bulge itself) is an X-structure. Even though it is possible to
include it as a separate component \citep{Savchenko2017, Smirnov2020}, this would add a considerable number of additional free
parameters to the model. In~this work we concentrate on the disk shape,
and hence, we fit the central concentration only to reduce its impact
on the disk component. In~the Results section we will mark galaxies where we detect the presence of an X-structure, and~where the obtained bulge parameters
suffer from this.

For a disk itself, we used two different
functions. The~first is the function that is described by the
Equation~(\ref{eq:simple_disk}), used to model disks without noticeable breaks. The~second is a function
(\ref{eq:broken_disk}) used to model broken disks. Both three-dimensional disk functions are included in the list of standard functions of the {\small IMFIT} package, and~during the decomposition process they are projected on a picture
plane via numerical line-of-sight integration. To~decide which of
the two disk models to apply to a particular galaxy image, we used the following approach: We started with decomposition using a simple unbroken disk and then inspected visually two-dimensional residual maps (data minus model)
and photometric cuts of the data and model made along the galaxy's major axis. If~a disk contains a break, its presence is
clearly visible during these inspections, and~approximate break parameters (break radius and break type) can also be
estimated alongside it. We then replaced the simple disk function with a broken disk in the configuration file, taking its
radial and vertical scales as initial conditions for a new model. This approach not only allows us to robustly detect
disk breaks, but~also to build a complex model with a broken disk using a simpler one as an intermediate step. The~procedure described above was applied to the $r$-band first, and~then these results were used as initial conditions for
the other~bands.

\begin{figure}[H]
 
  \hspace{-5pt}\begin{minipage}{0.32\linewidth}
    \centering
    \footnotesize PGC~90713\\
    \includegraphics[width=\linewidth]{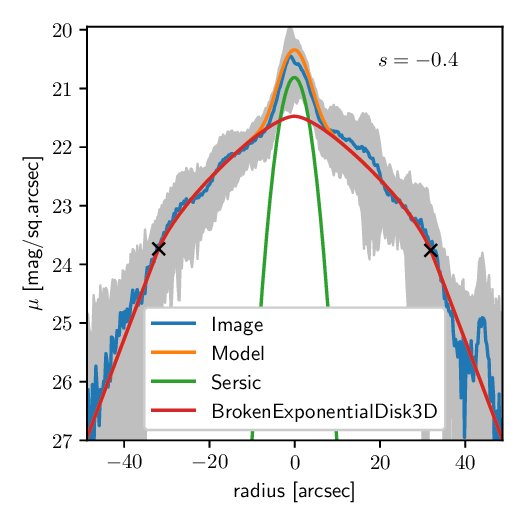}
  \end{minipage}
  \hfill
  \begin{minipage}{0.32\linewidth}
    \centering
    \footnotesize PGC~16144\\
    \includegraphics[width=\linewidth]{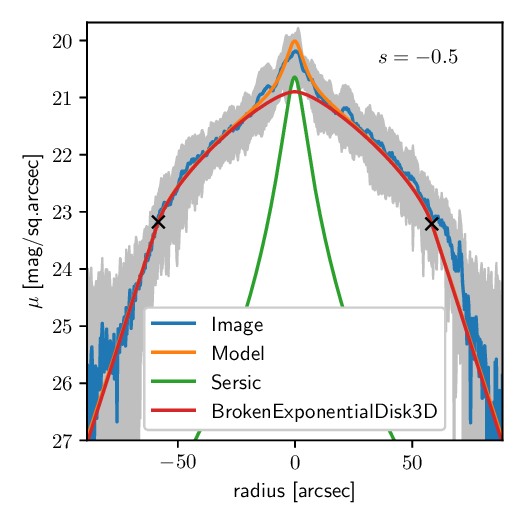}
  \end{minipage}
  \hfill
  \begin{minipage}{0.32\linewidth}
    \centering
    \footnotesize PGC~49296\\
    \includegraphics[width=\linewidth]{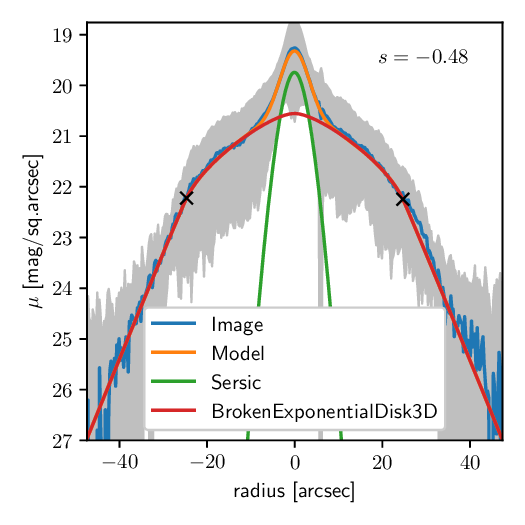}
  \end{minipage}

  \caption{Decomposition results for three galaxies: PGC~90713, PGC~16144, and~PGC~49296. Each panel shows photometric
    slices taken along the galaxy's major axis. Break locations are shown by x-marks. Numbers in top right indicate the
    break strength (see Equation~(\ref{eq:strength})). To~reduce the noise, for~each slice we compute the arithmetic
    mean of surface brightness inside a strip with a width of 1.6 arc seconds (roughly the PSF size). To~exclude the
    dust-attenuated regions, we take two parallel slices above~and below the dust lane and compute their
    average. The~blue line shows the observed brightness distribution, orange represents a full model, while green and
    red show bulge and disk models respectively. The~shaded region shows the standard deviation of the observed surface
    brightness inside a 1.6 arc-second strip.}
  \label{fig:decomposition_slices}
\end{figure}

\begin{figure}[H]
  \includegraphics[width=\linewidth]{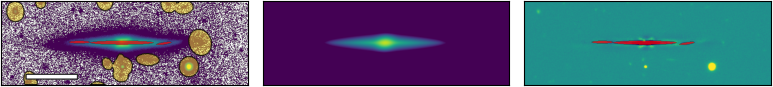}
  \includegraphics[width=\linewidth]{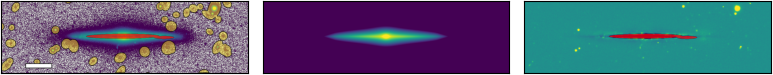}
  \includegraphics[width=\linewidth]{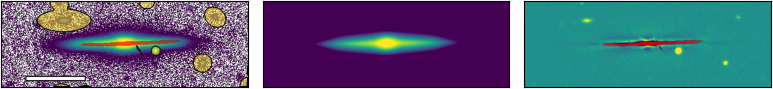}
  \caption{Decomposition results for the same three galaxies as in Figure~\ref{fig:decomposition_slices}, but~shown as
    2D image maps. Left column: original images in the $r$-band, with~yellow regions indicating the background objects
    mask; red regions---the manual dust mask. The white line in the bottom left marks a scale of 30''. Middle panels:
    model image. Right panels: ``image$-$model'', with~the dust mask overlaid.}
  \label{fig:decomposition_2d}
\end{figure}

Figure~\ref{fig:decomposition_slices} demonstrates the results of the decomposition for three galaxies: PGC~90713,
PGC~16144, and~PGC~49296. The~figure shows horizontal slices taken along major axes: blue---the observed brightness
distribution, green---the S{\'e}rsic function (bulge), red---the BrokenExponentialDisk3D function (disk), and~orange---the total model. To~exclude dust-contaminated regions (which were masked out), the~figure shows the average of two
slices taken above and below the dust lane. Figure~\ref{fig:decomposition_2d} shows the decomposition results for the
same three galaxies in the form of 2D image maps. The~left panels for each galaxy show the original images in the
$r$-band, with yellow regions indicating parts of the background that the objects mask, red regions representing the manual dust mask, and~the white line
indicating a scale of 30 arc seconds.
 The~middle panels show the model images, while the right panels show residual
``image$-$model'' maps with the dust mask~overlaid.

\section{Results}
\label{sec:results}

\begin{figure}[h]
  \centering
  \includegraphics[width=0.95\linewidth]{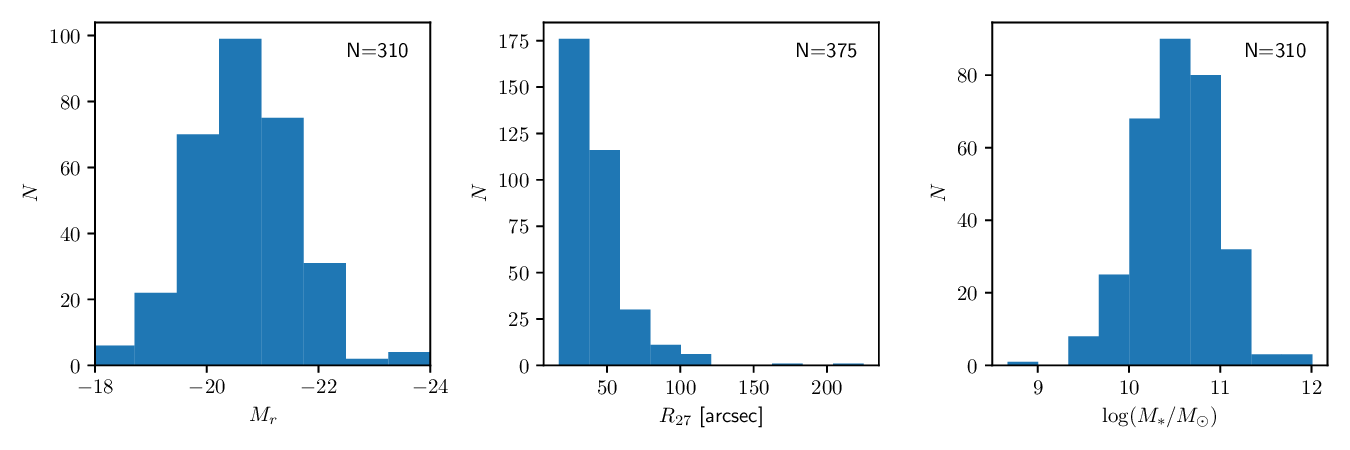}
  \caption{Distribution of sample galaxies by absolute magnitude in the $r$-band (left), radius of 27th-magnitude
    isophote in the $r$-band (middle), and stellar mass (right). Numbers of galaxies for which the corresponding
    parameter was obtained are shown in the top right corners.}
  \label{fig:sample_overview}
\end{figure}

Figure~\ref{fig:sample_overview} shows the distributions of the galaxies in our sample according to several parameters
inferred from the decomposition. The~left panel shows the distribution of absolute magnitudes in the $r$-band, which
were computed from the integrated magnitudes of bulge and disk models, and~luminosity distances to galaxies according to
the NED \url{https://ned.ipac.caltech.edu/} (Galactic extinction was corrected following \citep{Schlafly2011}). Since
only 310 galaxies in our sample have distance measurements in the NED, absolute magnitudes and stellar masses are
computed only for this subsample, and the provided distributions may not reflect the true distribution of a whole
sample. The~middle panel shows the distribution of our sample galaxies according to the radius of the 27th-magnitude
isophote in the $r$-band. This parameter was computed from the disk models obtained during the decomposition (we
neglected the bulges, since surface brightness in the outer regions is dominated by the disks). The~right panel shows
the distribution of galaxies by stellar mass computed from absolute magnitudes in the $g$- and $r$-bands using the
calibration from \citep{Ebrova2025}. It can be seen from this panel that the galaxies in our sample have stellar masses
from $\sim$10\textsuperscript{9.5} to \mbox{$\sim$10\textsuperscript{11.5} $M_\odot$}, with~a peak at
$10^{10.5}M_\odot$; thus they cover a entire mass range characteristic of disk galaxies, and~other works on disk breaks
often cover a similar mass rage: 10\textsuperscript{9}--10\textsuperscript{11} $M_\odot$ in~\cite{Tang2020},
10\textsuperscript{9}--10\textsuperscript{11.5} $M_\odot$ in ~\cite{MendezAbreu2017}, $M_\ast>10^{9.5}$
in~\cite{Erwin2012}. We note that our mass estimates were obtained using magnitudes from the decomposition, in~which we
at least partially reduced the impact of dust attenuation by applying dust masks. Dust attenuation is most significant
for edge-on galaxies, so our optical fluxes (and hence, the derived stellar masses) should be higher than those obtained
from simple aperture photometry. Nevertheless, a~small number of the most massive galaxies in our sample (with stellar
masses close to $10^{12}M_\odot$) could be outliers due to natural scatter in the measurements and possible systematics
in the flux-to-mass~calibration.

\vspace{-2pt}

\begin{figure}[h]
  \centering
  \includegraphics[width=0.95\linewidth]{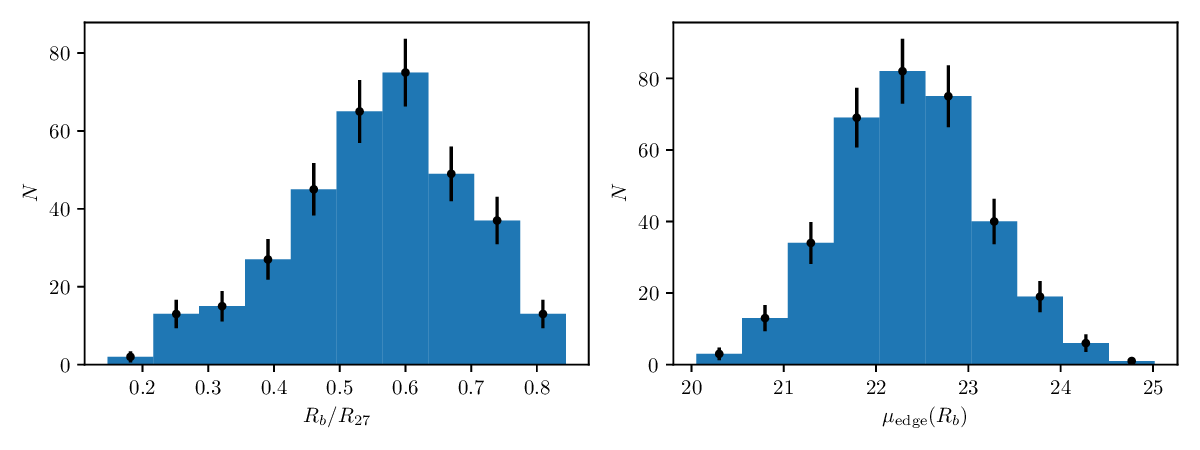}
  \caption{Left: Distribution of the sample break radii normalized by the disk's 27-th magnitude isophote in the
    $r$-band. Right: Distribution of breaks by the surface brightnesses at the break point in the $r$-band. Error bars show Poisson $\sqrt{N}$ uncertainty.}
  \label{fig:mu_rb_distribution}
\end{figure}

\subsection{Fraction of Breaks and Comparison to Previous~Works}
Out of the total of 375 galaxies in our sample, 33 galaxies ($8.8^{+1.6}_{-1.4}\%$) show no radial breaks (Type~I), 341
galaxies ($90.9^{+1.4}_{-1.6}\%$) exhibit downbending (Type~II breaks), and~only one galaxy ($0.3^{+0.4}_{-0.2}\%$) has
an upbending (Type~III) break (the uncertainties are $1\sigma$ Wilson intervals that were computed using
\textsc{proportion\_confint} function of \textsc{statsmodels.stats.proportion} Python package).
Figure~\ref{fig:mu_rb_distribution} shows the distributions of the breaks detected in our sample: left panel---break
radii normalized by the disk’s 27th magnitude isophote, and~right panel---surface brightness at the break point (both in
the $r$-band).

The fraction of galaxies with breaks in our sample is considerably higher than in many previous studies, where most
estimates find that roughly half of the galaxies host Type~II breaks \citep{Tang2020, Erwin2008, Gutierrez2011},
with~some reporting up to 82\% \citep{MartinNavarro2012}. In the following subsections we discuss possible reasons for
this discrepancy: impact of the photometric depth, possible disk flaring and an interplay between two disks (thin and
thick) with different radial scales.

\subsubsection{Impact of Photometric Depth and Disk~Orientation}
The DESI Legacy DR10 survey used in this work provides considerably deeper imaging compared to many older
wide-field surveys. For~example, the~Sloan Digital Sky Survey, which is often used to investigate breaks in disk
galaxies, is almost two magnitudes shallower in the $g$-band: our estimate of its $3\sigma$ depth made in $10\times 10$
arc-second squares is $27.17 \pm 0.13$ magnitudes per square arc second compared to $28.98 \pm 0.41$ for the DESI
Legacy~DR10.

\begin{figure}[h]
  \centering
  \includegraphics[width=0.5\linewidth]{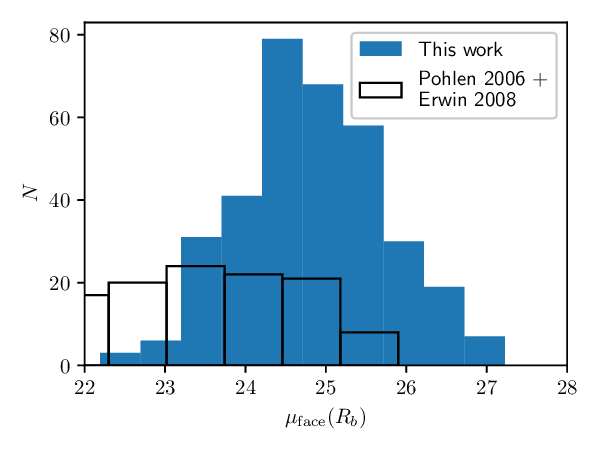}
  \caption{Distribution of surface brightness at the break point obtained in this work and recomputed to a face-on
    orientation (filled blue), compared with measurements from \citep{Pohlen2006, Erwin2008} (unfilled black).}
  \label{fig:mu_rb_distribution_fon}
\end{figure}

A second reason for the higher fraction of breaks detected in this work is that we investigate a sample of edge-on
galaxies. As~discussed in Section~\ref{sec:breaks_fon_eon}, projection effects lead to higher observed surface
brightness in edge-on galaxies compared to face-on or mildly inclined ones, and~the discrepancy can amount to several
magnitudes. This means that more distant regions of galaxies can be probed, where outer breaks may exist. To~check this
possibility, we recalculate the observed surface brightness at the break point that we obtained (shown on the right
panel of Figure~\ref{fig:mu_rb_distribution}) into a face-on orientation using
Equation~(\ref{eq:sb_diff_in_break}). This equation can be applied to our galaxies since we also obtained vertical
scales of their disks during the decomposition, a~quantity necessary for such a computation, which can be obtained
robustly only for edge-on galaxies. The~resulting distribution is shown in Figure~\ref{fig:mu_rb_distribution_fon}
(blue-filled histogram). One can see that the entire distribution has shifted considerably toward lower surface
brightness levels, so if the same galaxies were observed face-on, some breaks would be located in faint outer regions,
and~would hence be difficult to detect. For~comparison, we also plot on the same figure the combined results from
\citep{Pohlen2006, Erwin2008}. In~these works face-on and mildly inclined galaxies were investigated---i.e.,~those for
which projection effects do not provide favorable conditions for detecting breaks in faint outer regions. We note,
however, that this comparison should be treated as only qualitative, since the galaxies in the samples of
\citep{Pohlen2006, Erwin2008} are not strictly face-on oriented, and~recalculation of their surface brightness would
also be needed, but~without the information on their vertical-scale lengths, such correction cannot be performed. On~the
other hand, from Figure~\ref{fig:r_break_brightness_on_inclination} it is clear that this correction is relatively small
for disk inclinations below 60 degrees, which were analyzed in these works. Another point is that the samples considered
in this work and in \citep{Pohlen2006, Erwin2008} may differ in distribution based on the environment, Hubble type,
stellar mass, and~other parameters, which are known to correlate with the fraction of disk breaks. For~example, galaxies
with lower surface brightness are more likely to be detected if observed edge-on. In~any case, from~this figure it can
be seen that the number of breaks detected in \citep{Pohlen2006, Erwin2008} drops rapidly for surface brightness levels
below $\sim 25$ magnitudes per square arc second, whereas we detect a considerable number of breaks in these
faint~regions.

Another interesting result from Figure~\ref{fig:mu_rb_distribution_fon} is that on the faint side of the plot the number
of breaks detected in \citep{Pohlen2006, Erwin2008} is, on~the contrary, much higher than we detect. This can also be
explained as an outcome of projection effects. Breaks in regions of high surface brightness are often related to the
outer Lindblad resonances of bars (such breaks are dubbed II.o-OLR in \citep{Erwin2008}), and~the edge-on orientation
disfavors their detections because they are smeared out by the projection of the outer disk (see
Figure~\ref{fig:eon_fon_surfbri_compar}) and are also likely to be greatly affected by the~dust.

\subsubsection{Impact of~Flaring}

As mentioned in~\cite{Borlaff2016}, disk flaring may lead to an apparent Type~II break in an
edge-on galaxy even if the same disk would appear to be Type~I when viewed face-on. This happens because
the flaring reduces the flux in the disk plane by scattering some fraction of stars to larger
heights. In~this section, we check if such an effect may affect our results. We start with the
equation of a flared (but not broken) disk:
\begin{equation}
    \rho(R, z) = \rho_0 \exp \left(-\frac{R}{h_r}\right) \exp \left(- \frac{\left|z\right|}{h_z(R)} \right) \frac{h_{z,0}}{h_z(R)}.
    \label{eq:disk_with_flaring}
\end{equation}

Here, $h_z (R)$ is a function that describes how the vertical scale depends on radius, and~$h_{z, 0}$ is a vertical
scale of the disk inside the flare onset radius. The~last term is needed to keep the total number of stars at a given
radius: flaring makes the disk thicker, not more massive (see corresponding equations in~\cite{LopezCorredoria2002,
  Borlaff2016}). By~combining this equation with the outer part of a broken disk from Equation~(\ref{eq:broken_disk}),
we can find the analytical form of $h_z (R)$, i.e.,~we can find how strong a flare should be to produce the observed
strength of a downbending break. If~we define break strength such that $h_2 = \alpha h_1$ (with $0 < \alpha < 1$ for a
Type~II break) and $x=R-R_b$ is the distance from the break position outwards, the~solution for $h_z (R)$ would be
\begin{equation}
h_z(x) = \exp \left( \frac{1-\alpha}{\alpha} \frac{x}{h_1} \right) h_{z, 0}
\label{eq:flaring}
\end{equation}
(see the full derivation of this equation in the Appendix~\ref{sec:appendix_flaring}). And~vice~versa, if~we
have flaring in the form of

$$
h_z (x) = \exp{\left (f\frac{x}{h_1}\right) h_{z, 0}},
$$
with $f$ describing the flaring strength, then the observed break strength would be
\begin{equation}
    \label{eq:flare_to_break_strength}
    \alpha = \frac{1}{1+f}.
\end{equation}

There are several immediate consequences immediately of Equation~(\ref{eq:flaring}). First of all, to produce an
exponential break in radial surface brightness as defined in Equation~(\ref{eq:broken_disk}),
an~\textit{exponential} flaring is needed. \textit{Linear} flaring, which is often reported in the literature~(e.g., \cite{deGrijs1997,
  Narayan2002, Saha2009, Kasparova2020}), cannot produce an \textit{exponential} radial break. The~second point is
that the required flaring strength strongly depends on the break strength. The~observations of exponential flare
strength for the galaxy, where such measurements are most extensive, do not show values larger than $f=0.5\dots0.6$
\citep{Hammersley2011, LopezCorredoria2014, Bovy2016, Mackereth2017}, so if we take a flare strength of $f=0.6$ as the
most extreme case, we can use Equation~(\ref{eq:flare_to_break_strength}) to find that the most extreme break would be
$h_2 = 0.625 h_1$, or, in~terms of logarithmic break strength (see Equation~(\ref{eq:strength})), $s=\log
(0.625)=-0.2$. This value is very close to the one obtained in a numerical experiment by~\cite{Borlaff2016}.

The final note we would like to discuss here is that the relation
between the flare strength and the apparent break strength is found for the 
worst-case scenario: one-dimensional decomposition of a photometric slice taken along the galactic plane
(as was studied in~\cite{Borlaff2016}). In~such a case, the~light that left the disk plane
due to flaring is completely missed during decomposition. But~it is not the case
for the two-dimensional decomposition performed in this work. When we perform two-dimensional
decomposition, the~{\small IMFIT} package that takes the whole image as input still captures
some fraction of the flared light above and below the disk plane. In~other words, the~flared
light is not missing completely in the case of two-dimensional decomposition. This effect should
lead to a lower measured break strength than the one predicted by
Equations~(\ref{eq:flaring}) and (\ref{eq:flare_to_break_strength}). To~check for this possibility, we
performed the following numerical experiment.

\begin{figure}
  \centering
   \includegraphics[width=0.6\linewidth]{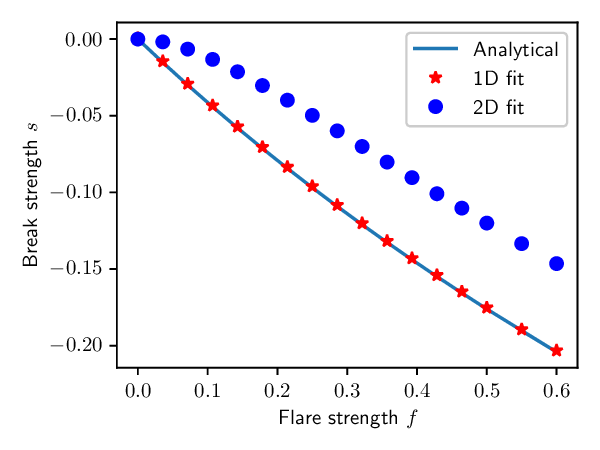}
   \caption{Results of the numerical experiment on the decomposition of a flaring disk by a broken disk model,
     highlighting how the measured break strength depends on the true value of the flare strength.  Solid line:
     analytical prediction from Equation~(\ref{eq:flare_to_break_strength}); asterisks: result of one-dimensional
     decomposition; dots: results of two-dimensional decomposition.}
  \label{fig:flare_vs_break}
\end{figure}

We added a function of a flared disk according to Equation~(\ref{eq:disk_with_flaring}) to the
{\small IMFIT} package following the package user guide. Then we created a set of model images
of flaring disks with flaring strength $f$ in the range 0 to 0.6. After~that we performed two types
of decomposition by a single broken disk: one-dimensional decomposition of a photometric
slice taken along the disk plane (as in \citep{Borlaff2016}) and two-dimensional decomposition
using {\small IMFIT}. Then we checked how the measured break strength depends on the input flare
strength. The~results of this experiment are shown in Figure~\ref{fig:flare_vs_break}, where the
measured break strength is plotted against the input flare strength. Here a solid line shows the
analytical prediction of Equation~(\ref{eq:flare_to_break_strength}), asterisks show results of
one-dimensional decomposition, and~dots show results of two-dimensional decomposition. As~expected,
the measured break strength increases with the flaring strength. It is evident that numerical 
results of one-dimensional decomposition align perfectly along the analytical prediction, which
validates our findings. We also note that in two-dimensional decomposition the measured break
strength is \textit{lower} than in the one-dimensional case, as~we expect, due to partial capturing
of flared light by a two-dimensional function.

The main result of this section is that even though stellar disk flaring can indeed produce a Type~II break appearance
in an edge-on disk, the~strength of this effect is not high enough to explain the observed distribution of Type~II
breaks.  Any reasonable flaring strength cannot produce a downbending break stronger than $s\approx-0.2$ during
one-dimensional decomposition, and~results in even weaker breaks ($s\lessapprox -0.15$) for two-dimensional
decomposition. In~fact, almost all breaks in galaxies of our sample are stronger than this limit (see
Section~\ref{sec:two_kinds_of_breaks}). Therefore we do not expect our galaxies to change their break type due to
flaring, and~rule out flaring as a possible source of the high fraction of Type~II disk breaks in our sample.

\subsubsection{{Thin and Thick~Disks}}
So far, we have carried out our analysis considering only a single disk. However, a~disk often consist of two
  components, thin and thick, with~different scale heights~\cite{Comeron2018}. If~their radial brightness distributions
  are similar, then a single disk model with free-fitted $n$ may still describe the combination of both of them
  reasonably well. More importantly for our study, in~this case we would correctly obtain the properties of disk break.

  However, the~thick and thin disks may have different radial scales. For~example, if~the thick disk (whose central
  brightness is fainter) has a larger radial scale than the thin component, at~the periphery it may overtake the thin
  disk in brightness, creating the appearance of a Type III radial break~\cite{Comeron2018}. Similarly, the~vertical
  disk profile at larger radii becomes thick-disk-dominated, yielding a higher vertical height and causing apparent
  flaring. Thus, fitting these two disks with a one-disk model may potentially bias the results regarding break
  detection.

  To examine cases like the one described above, we fit two disks---a thin and a thick one---to the
  image. Examining images, models, and~2D residual maps, we select nine galaxies that likely have the most
  prominent thick disks and 10 other random galaxies, and~fit them with a model containing a bulge and two disks
  without breaks. Note that our aim is not to obtain a perfect two-disk model (hence, we do not add breaks) but
  rather to ensure that our conclusion regarding break statistics is correct.

\begin{figure}[h]
  \centering
  \includegraphics[width=0.9\linewidth]{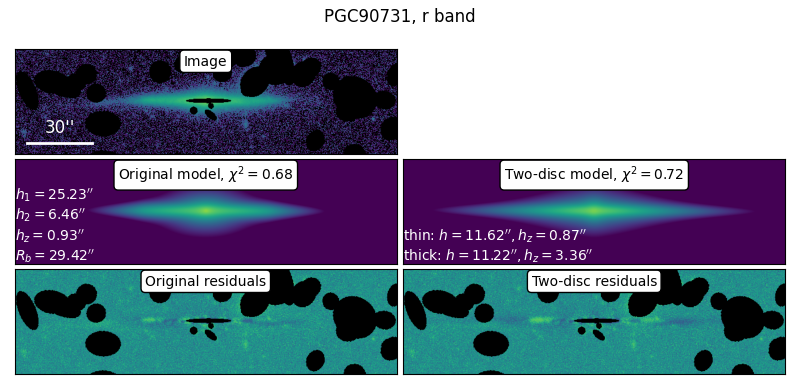}
  \caption{A comparison between our original model containing a single disk with a break (left column) and a model with
    two disks, one thick and one thin (right column). From~top to bottom: image, model, and~residual image. Disk
    parameters are listed at the corresponding model images.}
  \label{fig:two_disc_comp}
\end{figure}

  We compare this two-disk model with the original model, with~an example shown in Figure~\ref{fig:two_disc_comp}. It
  turns out that, both for the subsample of galaxies with an apparently prominent thick disk and for those without it,
  the~two-disk model almost always has worse fit quality: the two-disk model usually fails to reproduce the appearance
  of Type~II profiles (which are observed in 16 out of 19 galaxies across the two subsamples). In~these 16 galaxies,
  only one has its fit quality improved when two disks are used (and it is a galaxy with a prominent thick
  disk). The~median $\chi^2$ for the 9 galaxies with an apparent thick disk is 1.88 for the original models and 2.31 for
  the two-disk models, and~for the control sample of 10 galaxies these values are 0.97 and 1.16, respectively.

Again, this analysis does not imply that galaxies in our sample lack thick disk components, but rather it shows that
Type~II profile appearance cannot be explained by the combination of two disks without breaks. Indeed, if~the thick disk
has the same or a smaller radial scale than the thin one and both have no breaks, the~radial profile would be Type~I,
defined only by the thin component. Otherwise, the Type~III profile could appear as described above. However, for~a
Type~II profile to appear, the~more prominent component of the disk has to demonstrate a genuine downbending
break. In~other words, the~observed fraction of Type~II profiles is not exaggerated due to the single-disk approach, nor
do we miss Type~III profiles with a given precision.

\subsubsection{{More Complex Disk Structures?}}
Above we investigated two possibilities that could lead to the erroneous attribution of an edge-on galaxy to a
  broken type: flaring and two-disk composition. In~this section we describe another test to check if some complicated
  dependence of vertical brightness distribution on radius could lead to the false appearance of a broken disk. The~idea
  is to project the vertical light distribution onto a disk plane and check if this projected light distribution still
  gives the broken radial profile. To~do so, we adopt the following approach: First, we rotate an image of a galaxy such
  that it appears horizontal. Then, for~each point in the galaxy plane, we compute the total flux from all pixels above
  and below this point (taking a stellar mask into account). As~a result, we obtain a projection of a galaxy onto an
  infinitely thin disk. In~this projection, we compensate for all irregularities of the vertical brightness distribution
  such as flaring, because~we combine the total light from all heights at any given radius regardless of the underlying
  vertical brightness distribution. If~we observe a radial break in this projected brightness distribution, then it is
  not an artifact of flaring, a~thin--thick disk dichotomy, or~any combination of such effects.

\begin{figure}[h]
  \centering
  \includegraphics[width=\linewidth]{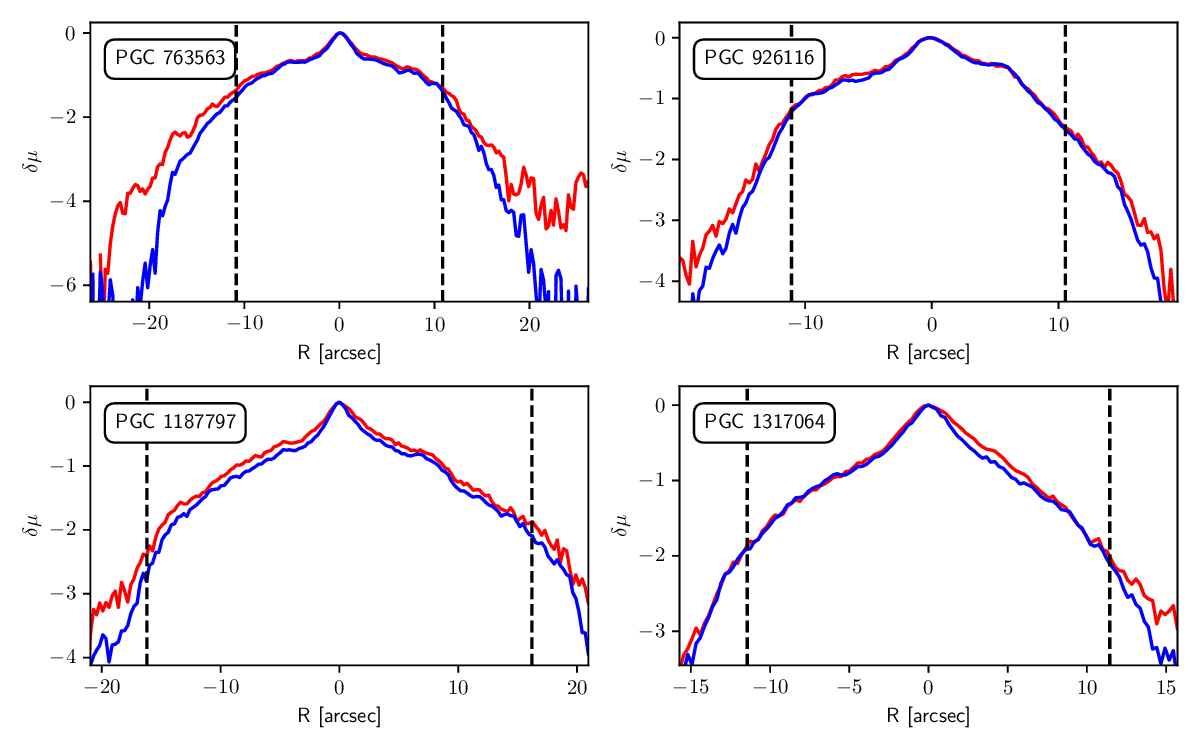}
  \caption{Results of the experiment on brightness projection onto a disk plane are shown for four galaxies with Type~II
    breaks. In~ each panel, the~blue line shows a photometric cut along the disk plane, and the red line shows radial
    brightness distribution of the projection. Both brightness distributions are shifted vertically such that their
    peaks match. Vertical axes show brightness fading in stellar magnitudes relative to the peak. Vertical dashed lines
    show break positions computed from 2D decomposition.}
  \label{fig:projected_disks}
\end{figure}

  The results of this experiment are presented in Figure~\ref{fig:projected_disks} for four galaxies of our sample that
  present Type~II breaks. For~each galaxy, the~blue line shows a regular photometric cut taken along the disk plane,
  and~the red line shows the brightness distribution projected onto the disk plane. Both curves are shifted vertically
  such that the central brightness of one matches that of the other. It is clear from this figure that for all four
  galaxies the difference between two brightness distributions increases at larger radii, which may indicate radial
  variations of $h_z (R)$ due to flaring/a thick disk: there is more light at higher altitudes in outer regions of
  galaxies. This is especially prominent for PGC~763563. But~for all galaxies the break persists even in the projected
  brightness distribution, so we can conclude that these galaxies indeed have downbending breaks. This is also true for
  all the other galaxies of our sample; we have not found any galaxies where the break disappears on the projected
  brightness distribution.  Finally, we note that in the projected brightness distribution the break strength may be
  lower because the disk's inner and outer radial scales and break radii depend on height \citep{Pohlen2007}, which
  leads to smearing of the projected profile.

\subsection{Outer Disk Scale vs. Surface Brightness at Break~Point}
\begin{figure}[h]
  \centering
  \includegraphics[width=\linewidth]{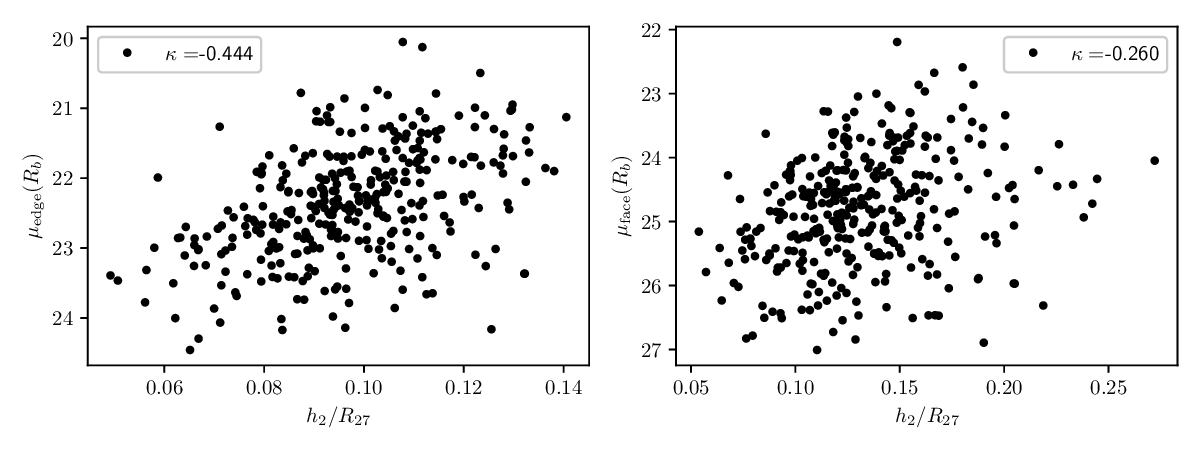}
  \caption{Radial exponential-scale length of an outer disk (normalized by the 27th magnitude isophote radius) vs.
    surface brightness at the break point (all values in r-band). Left-panel: observed values, right panel: recalculated
    to the face-on orientation.}
  \label{fig:outer_scale_vs_sb}
\end{figure}

As noted in the discussion of Equation~(\ref{eq:sb_diff_in_break}), when a galaxy is observed edge-on, there should be
an artificial projection-induced correlation between the radial-scale length of the outer disk and the observed surface
brightness at the break point. This correlation should appear only close to the edge-on view and is due to projection
effects, without~reflecting a real correlation of intrinsic galactic properties. Figure~\ref{fig:outer_scale_vs_sb}
demonstrates this effect: its left panel shows the $r$-band surface brightness at the break point as a function of the
radial-scale length of the outer disk (normalized to the radius of the 27th-magnitude isophote for consistency). Indeed,
there is a rather strong correlation (Pearson's correlation coefficient is $-0.444$): galaxies with a more extended
outer disk have higher break point surface brightness. The~right panel of this figure shows the same parameters
recomputed to the face-on orientation. In~this case, the~correlation coefficient drops significantly (to $-0.260$). This
means that the observed correlation is a result of projection effects and that one must be careful when analyzing
correlations between decomposition parameters for galaxies with different~orientations.

\subsection{Two Kinds of Downbending~Breaks}
\label{sec:two_kinds_of_breaks}
\begin{figure}[h]
  \centering
  \includegraphics[width=\linewidth]{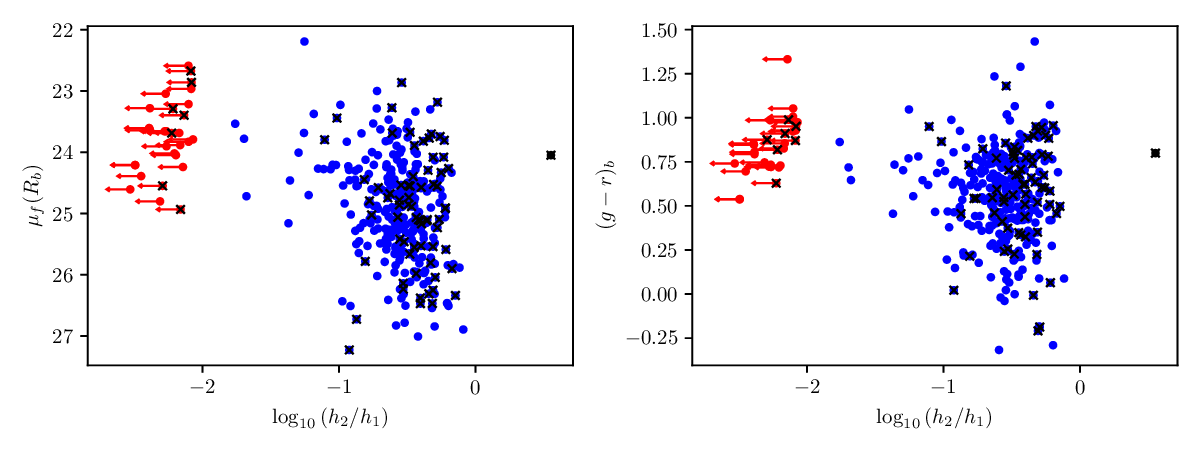}
  \caption{Left panel: face-on surface brightness for break point vs. break strength. Right panel: $(g-r)$ color for
    break point vs. break strength. Red dots mark disks with strength $< -2.0$, and blue dots mark those with strength
    $>-2.0$. The~black dot is a sole galaxy with a Type~III break (PGC~731, see Appendix~\ref{ap:pgc_731} for details
    about this object). Crosses mark galaxies that host apparent X-structures.}
  \label{fig:two_kinds_of_breaks}
\end{figure}

Figure~\ref{fig:two_kinds_of_breaks} shows how the surface brightness at the break point (recomputed to the face-on
orientation) and the disk color at the break point depend on the break strength. One thing that is readily seen from
this figure is that there are two distinct sets of points: very strong Type~II breaks (with the strength values
$<-2.0$, marked as red dots) and rather mild Type~II breaks (with the strength values $\gtrsim -1.5$, marked as blue
dots). There are almost no breaks between these two~groups.

\begin{figure}[h]
  \hspace{-5.5pt} \begin{minipage}{0.32\linewidth}
    \centering
    \footnotesize PGC~90630\\
    \includegraphics[width=\linewidth]{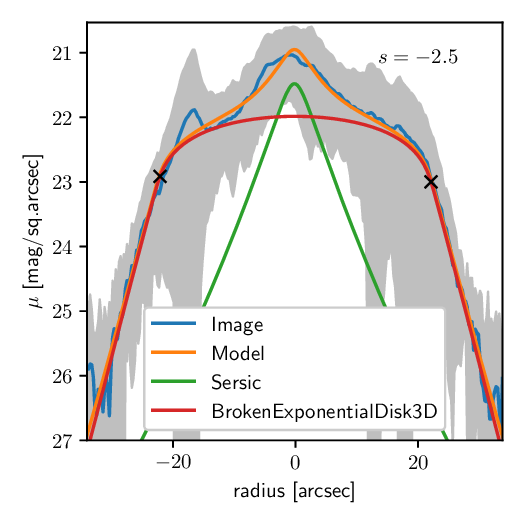}
  \end{minipage}
  \hfill
  \begin{minipage}{0.32\linewidth}
    \centering
    \footnotesize PGC~1148678\\
    \includegraphics[width=\linewidth]{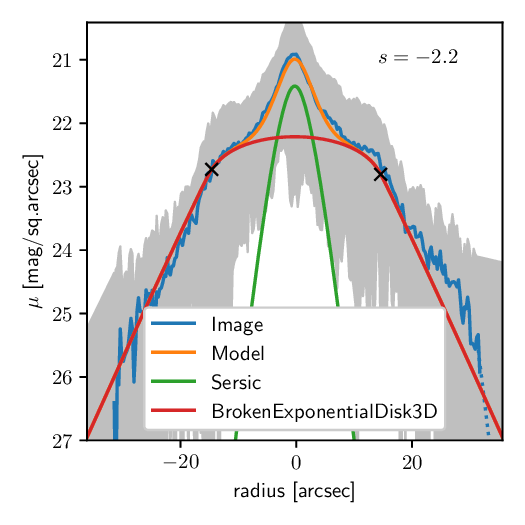}
  \end{minipage}
  \hfill
  \begin{minipage}{0.32\linewidth}
    \centering
    \footnotesize PGC~2160344\\
    \includegraphics[width=\linewidth]{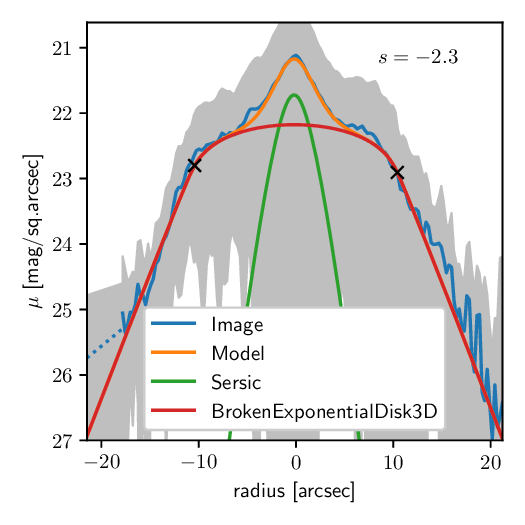}
  \end{minipage}

  \caption{Decomposition results for three galaxies with very strong breaks. All markings are the same as in
    Figure~\ref{fig:decomposition_slices}.}
  \label{fig:decomposition_slices_strong_breaks}
\end{figure}

In all cases, the very high negative values of the break strength parameter are due to large radial-scale lengths of the
inner disk. Figure~\ref{fig:decomposition_slices_strong_breaks} demonstrates examples of three such galaxies. It is
remarkable how the true break locations (shown by x-marks in the figure) differ from the apparent ones due to projection
effects (see Figure~\ref{fig:eon_fon_surfbri_compar} and the related discussion in text). Some disks exhibit an almost
flat light distribution in their inner regions, so very high values of $h_1$ are required to fit it. Interestingly,
studies of face-on and mildly inclined galaxies usually do not demonstrate galaxies with such high break
strengths. For~example in \citep{Maltby2012} the strongest downbending break has $s\approx-0.8$, while in
\citep{Laine2014} the most extreme Type~II break has a strength of about $-1$ (such galaxies are dubbed ``flat inner
disks'' in that work). However, the works that focus on edge-on galaxies do show some galaxies with very strong breaks
in their samples, although~the number of articles with large samples of edge-on galaxies is low. For~example,
in~\citep{Comeron2012} in a sample of 70 edge-on galaxies there is a galaxy with $s=-1.4$
(ESO~469-015). In~\citep{Smirnov2026} the decomposition of a sample of 71 edge-on galaxies is performed, and the authors
note that for some of them the radial-scale length of the inner disk reaches such high values that the light
distribution in the inner regions can be effectively considered flat. We note here that in this case the model
  degenerates in the $h_1$ value: if $h_1 \gg r_b$, the surface brightness of the inner disk becomes almost flat, and
  further increasing the $h_1$ value does not lead to significant changes in the model, and~therefore in the obtained
  $\chi^2$ value. This means that the true values of $h_1$ and $s$ become undefined in such models. To~emphasize this,
  we marked strong breaks as upper limits in Figure~\ref{fig:two_kinds_of_breaks}. One should treat these values
  carefully, and~only use them as an indicator of a strong break in the galaxy.  \vspace{2pt}

\begin{figure}[h]
  \includegraphics[width=\linewidth]{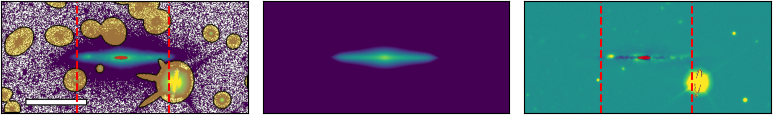}
  \includegraphics[width=\linewidth]{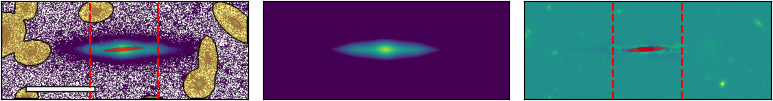}
  \includegraphics[width=\linewidth]{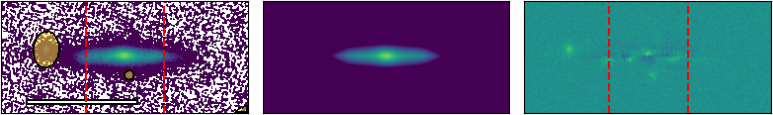}
  \caption{Decomposition results for the same three galaxies as in Figure~\ref{fig:decomposition_slices_strong_breaks},
    shown as 2D image maps. All markings are the same as in Figure~\ref{fig:decomposition_2d}, with~one exception:
    vertical dashed red lines indicate break locations.}
  \label{fig:decomposition_2d_strong_breaks}
\end{figure}

One possible explanation for a flat inner disk is the presence of an additional component such as an outer ring (see
\citep{Laine2014}), which adds extra light that, when combined with an exponential disk, may result in an almost flat
light distribution. Another possibility is a strong bar which can either manifest itself as an additional component,
or~form a ring at the outer Lindblad resonance (OLR) which leads to an apparent Type~II break~\cite{Pohlen2006}. In the
face-on orientation, such components would be easily detectable, but~in the edge-on view their identification is more
complicated due to projection effects. Another possibility is strong dust attenuation that absorbs the light from
central regions. To~check for clues, we examine 2D plots of original images, models and residuals for some galaxies with
strong breaks in Figure~\ref{fig:decomposition_2d_strong_breaks}, with~break locations marked. It can be seen from these
images that there are no obvious structures located near the break radii that could cause a flat inner disk
appearance. Even the residual maps, which usually highlight substructures that are not accounted for by the model, do
not show anything significant near or inside the break locations. In~particular, we do not see the presence of
X-structures in the residual maps, features that are manifestations of a bar's presence in the edge-on orientation
\citep{Combes1981, Raha1991}. A~strong bar could result in a density depression in the inner disk regions and even lead
to disks with central holes (so-called antitruncated disks) \citep{Smirnov2020}. We checked residual maps of all
  galaxies for the presence of X-structures using \citep{2022MNRAS.512.1371M}, and~found that the number of galaxies
  that host X-structures is 69 out of 311 galaxies (22\%) and 7 out of 30 galaxies (23\%) for weak and strong breaks
  respectively, so the presence of bars/X-structures alone cannot explain a strong break. It can also be seen that
these galaxies are not exceptionally dusty ones (at least, the~dust lanes are not prominent in the optical images).

As a final note, we would like to point out that from the radial slices alone
(Figure~\ref{fig:decomposition_slices_strong_breaks}) it may appear that there is a degeneracy between the bulge and the
inner disk: if the bulge becomes fainter, the~inner disk will become steeper to compensate for it, and~this will
resolve the flat inner disk problem. However, since we employ the two-dimensional decomposition, this is not the case. If~we reduced the bulge brightness, this would be readily noticeable in the 2D residual maps as two bright regions above
and below the disk center, where the bulge light is underestimated and where the disk could not account for it. There is
no difference between the mean bulge-to-total ratios between galaxies with strong and weak breaks ($0.23\pm 0.11$ and
$0.22 \pm 0.13$ in the $r$-band respectively), whereas in the case where the bulge becomes larger to account for some
disk light and produce a flat disk appearance, we would expect a systematic difference between the two subsamples. There
is also no difference in the ratio of bulge effective radius to disk vertical scale ($2.67 \pm 1.17$ and $2.50 \pm 1.48$
in the $r$-band accordingly), so we rule out the possibility that galaxies with strong breaks host geometrically smaller
bulges that would make it easier for them to blend in with the disks. We also do not find a significant difference in
the mean stellar mass between galaxies hosting weak and strong breaks ($\log (M_\ast/M_\odot)=10.52 \pm 0.46$ and
$10.54 \pm 0.36$).

\begin{figure}[h]
  \centering
  \includegraphics[width=\linewidth]{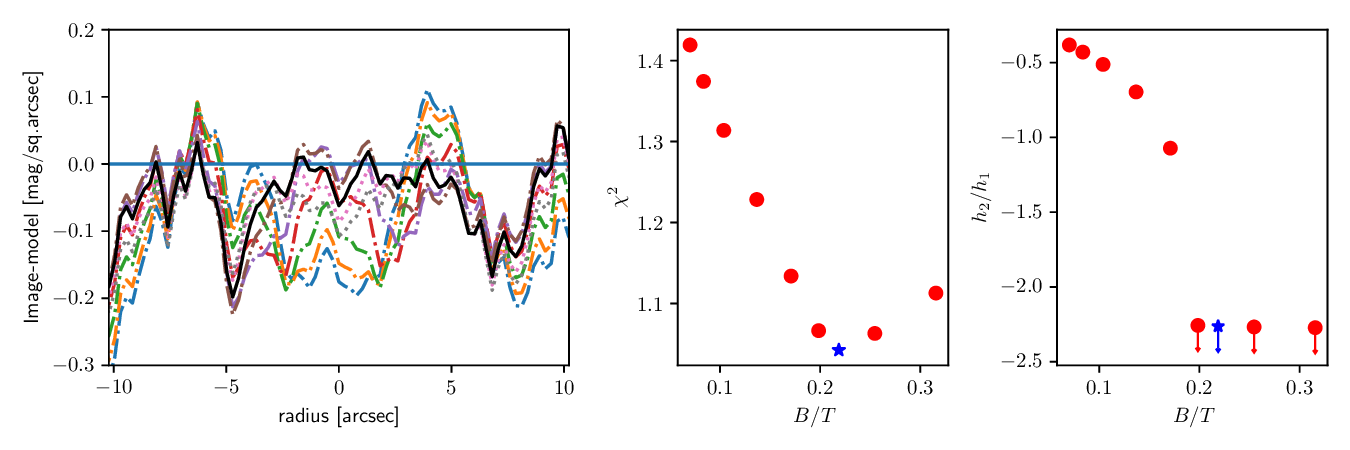}
  \caption{Results of a test for the degeneracy between bulge parameters and the break strength (see text). Left panel:
    photometric cuts along the disk plane in the ``image-model'' residual map for a set of decompositions with different
    values of bulge-to-total ratio. The solid black line shows the results of the best decomposition, dotted lines
    represent B/T fixed to larger values, and dot-dashed lines represent B/T fixed to lower values. Middle and right
    panels---$\chi^2$ and break strength values as a function of B/T value, with blue stars showing optimal
    decomposition with free bulge parameters.}
  \label{fig:degeneracy_expirement}
\end{figure}

To investigate the possibility of degeneracy between the bulge parameters and the inner disk scale we performed
  the following experiment: For~a PGC~2160344 galaxy (a strong break without prominent X-structure) we ran a set of
  decompositions where we forced the bulge to converge to brighter/fainter bulge-to-total ratios than those of the best
  fit. It is natural to expect that if we force the bulge to have a lower bulge-to-total ratio, the~inner disk scale
  will become larger to compensate for the missing bulge light. The~question is whether this results in the same quality
  fit (i.e., whether there is true degeneracy between bulge parameters and the $h_1$
  value). Figure~\ref{fig:degeneracy_expirement} demonstrates the results of such an experiment. On~the left panel we
  show photometric cuts taken from an ``image-model'' residual map parallel to the disk plane outside the dust lane
  (only central regions inside the break radius are shown).  The black solid line shows the result for a best-fit
  decomposition with free bulge parameters. Dotted lines show residuals of decompositions where we forced the bulge to
  have larger B/T ratios, and~dot-dashed lines show residuals where we forced the bulge to have lower B/T ratios. It is
  clear from this plot that models with both brighter and fainter bulges demonstrate larger residuals. The~same result
  but in terms of $\chi^2$ values is shown in the middle panel of the same figure. Here the blue star shows the
  best-fitting $\chi^2$ value, and~red dots show results with fainter/brighter bulges, all of which have larger
  $\chi^2$. The~right panel shows the break strength value as a function of the fixed bulge-to-total value. It is clear
  that if we force the bulge to have a lower value, we can obtain a significantly lower break strength, but~at the cost
  of a worse fit. The~Bayesian information criterion (BIC) value of the best fit is 28,721 compared to 31,224 for the
  one with a fainter bulge but lower break strength (the one with $h_2/h_1\approx-1$ in the figure), which makes the
  best fit with a free bulge the preferable one.

A possible solution to this problem could be obtained by performing the decomposition of highly inclined---but not
edge-on---galaxies, where the dust impact, bulge-disk degeneracy and projection effects are lower.

\subsection{Disk Color vs. Break~Strength}
Since we performed photometric decomposition of the galaxy images, we obtained fluxes for individual components (bulge
and disk) separately. Performing the decomposition in different passbands allows us to obtain the colors of these
components. To~do so, we converted the integrated fluxes of the components into magnitudes and accounted for the
Galactic dust extinction in all photometric bands, with calculations performed following recalibrated SFD dust maps
\citep{Schlafly2011} (the corrections were calculated from $E (B-V)$ reddening for $R_V=3.1$ using coefficients from
their Table~6)

\begin{figure}[h]
  \centering
  \includegraphics[width=0.65\linewidth]{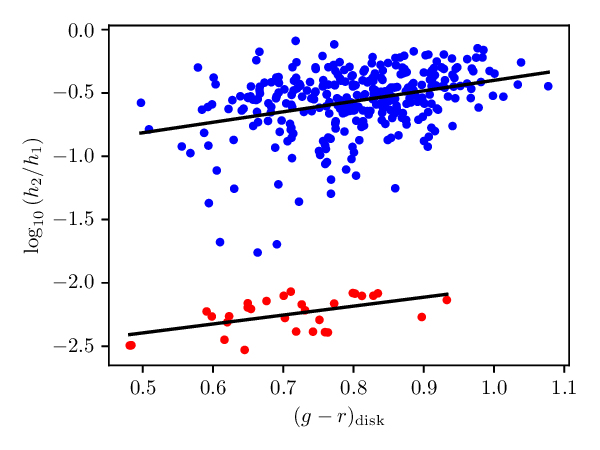}
  \caption{Break strength in the $r$-band vs. disk $(g-r)$ color. Blue dots: weak breaks (strength $>-2$), red dots:
    strong breaks (strength $<-2$). Solid black lines: linear regressions for the two break groups.}
  \label{fig:color_break_strength}
\end{figure}

Figure~\ref{fig:color_break_strength} shows the relation between the integrated disk $(g-r)$ color and the break
strength. As~in the previous subsection, we divided all the breaks into two subsamples: strong ones (with break strength
$<-2$) and weak ones (with strength $>-2$). As~can be seen from the figure, both groups of breaks exhibit significant
correlations. For~weak breaks, Pearson's correlation coefficient is $0.337$ (\emph{p}-value $1.06\cdot 10^{-9}$), while
for strong breaks, the correlation coefficient is $0.539$ (\emph{p}-value $0.002$). We performed linear regression on
the data points of both break groups and obtained the following results: for weak breaks
$s=0.828\cdot (g-r) _{\mathrm{disk}}-1.228$, and~for strong breaks $s=0.705\cdot (g-r) _{\mathrm{disk}}-2.747$. It is
notable that two such distinct groups of breaks follow a ``color--strength'' relation with very similar slopes. We also
note, that disks with strong breaks are on average somewhat bluer than disks with weak breaks, but~their color ranges
overlap greatly, so it is impossible to distinguish two types of disks based on color~alone.

The interpretation of this correlation between the disk color and the break strength can be made within the framework of
models that invoke star-formation to explain the generation of breaks. Bluer disks should host more intense current or
recent star-formation and therefore should be more prone to forming prominent breaks. As~mentioned in the Introduction,
many studies demonstrate, that breaks are observed more frequently in galaxies of later Hubble types, where
star-formation is higher \citep{Pohlen2006, Erwin2008, Gutierrez2011, Laine2016, MendezAbreu2017, Tang2020}, whereas
when a galactic star-formation is suppressed, for~example, by~ram pressure stripping, the number of breaks decreases
\citep{Erwin2012, Head2015, Pranger2017, Pfeffer2022, Mondelin2025}. The~correlation we report here is therefore in line
with these~findings.

\subsection{Break Strength as a Function of~Wavelength}
In \citep{Bakos2012} on a small sample of face-on to intermediately inclined galaxies from SDSS Stripe~82
\citep{Jiang2014}, it was demonstrated that the strength of Type~II breaks decreases, on~average, with~increasing
wavelength in the optical bands from $u$ to $z$. This result was confirmed by~\citep{Laine2016} using a larger sample of
mildly inclined galaxies from the S$^4$G sample \citep{Sheth2010}: $u$-band breaks turned out to be stronger than those
measured in the~infrared spectrum.

\begin{figure}[h]
  \centering
  \includegraphics[width=0.6\linewidth]{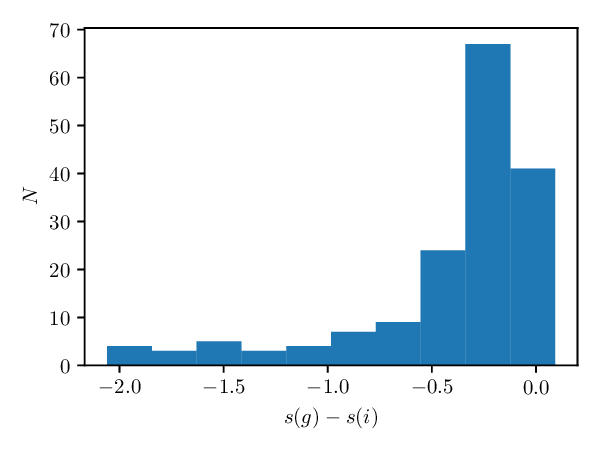}
  \caption{Distribution if differences between breaks strengths measured in $g$ and $i$-bands. For~Type~II breaks, lower
  values mean stronger breaks in the $g$-band.}
  \label{fig:strength_different_bands}
\end{figure}

Here we test this result for a sample of edge-on galaxies. Figure~\ref{fig:strength_different_bands} shows the
distribution of differences between break strengths measured in the $g$- and $i$-bands. Since the footprint of the DESI
Legacy survey in the $i$-band is smaller than that in the $g$-band, only 165 galaxies in our sample have images in both
bands. In~this figure we also exclude the sole upbending galaxy in our sample, so only Type~II galaxies are presented;
thus, lower values of $s (g)-s (i)$ indicate stronger breaks in the $g$-band that in the $i$-band. As~can be seen from the
figure, all galaxies show negative values of $s (g)-s (i)$, meaning they have stronger breaks at bluer wavelengths. This
confirms the result from \citep{Bakos2012,Laine2016} obtained with a different method and for a different sample of
galaxies.

\vspace{-3pt}

As in \citep{Laine2016}, we found that the main source of the dependence of break strength on wavelength is
related to the changes in the inner disk radial scale rather than the outer one. We computed the normalized average
differences between the radial scales of inner and outer disks in the $g$- and $i$-bands as
$$
\Delta h_i = \frac{h (g)-h (i)}{h (g)}, \quad i=1, 2,
$$
and found that $\Delta h_1=0.44\pm 0.27$, while $\Delta h_2=0.03\pm 0.13$. All inner disks have larger radial scales in
the $g$-band than in the $i$-band. The~differences in radial scales for the outer disks are distributed around zero with
considerably lower scatter. If~the break is caused by a drop in star-formation and the outer disk is populated by radial
migration of stars, we should observe larger radial scales of outer disks at longer wavelengths, since older stellar
populations have had more time to move outward. The~same systematic behavior should be observed if a break appears as a
result of a local enhancement in sSFR both inside and outside the break radius, as~was seen in simulations by
\citep{Chen2026}.  In~this work, we do not observe a statistically significant dependence of the outer disk scale on
the~wavelength.

We should mention that although we attempted to reduce the influence of dust on the decomposition results by masking
dusty regions, non-planar dust can still affect the measured parameters as a function of wavelength. For~more robust
results, decomposition of a comparably sized sample using a radiative transfer approach is needed to account for the
dust more carefully. However, such an analysis is yet to be~done.

\subsection{Rotation~Velocity}
Although galaxies in our sample are relatively close, only a minority of them (\mbox{100 objects}) have maximal rotation
velocity estimates in the HyperLEDA database \citep{Makarov2014}. Stellar masses derived from our photometry
measurements obtained from decomposition models correlate well with $V_{\mathrm{max}}$ (Pearson's correlation 0.781,
\emph{p}-value $3\cdot 10^{-22}$).

\begin{figure}[h]
  \centering
  \includegraphics[width=0.85\linewidth]{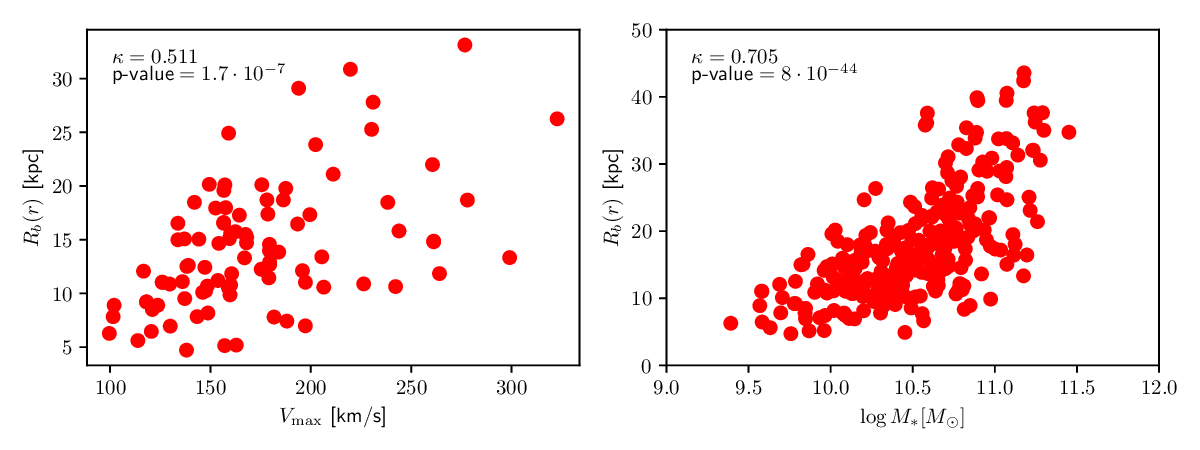}
  \caption{Break radius in $r$-band (left) and stellar mass (right) vs. maximum rotation velocity.}
  \label{fig:vrot_rbreak}
\end{figure}

We found that there is a significant correlation between the break radius expressed in absolute units and the maximum of
the rotation velocity.  Pearson's correlation coefficient is 0.511, and~the \emph{p}-value is $1.7 \cdot 10^{-7}$ (see
Figure~\ref{fig:vrot_rbreak}, left panel).  The correlation between the break radius and the rotation velocity was
studied in several articles because some theories associate breaks with maximum angular momentum in disks. It seems that
when the break radius is expressed in relative units, the~correlation is weak or absent. In~\cite{Pohlen2006,
  vanderKruit2008} the break radius was expressed in terms of inner disk radial scales and did not show a significant
correlation with rotation velocity. In~\cite{Gutierrez2011}, where the correlation was found, the~break was expressed in
units of $R_{25}$. In~\cite{MartinNavarro2012} a strong correlation between $R_b$ and $V_{\mathrm{max}}$ was found if
the break radius is expressed in absolute units. We confirm these findings: according to our data there is a significant
correlation if $R_b$ is expressed in parsecs, whereas if we normalize it by $R_{27}$, the~correlation disappears.  Since
the stellar mass derived from decomposition correlates well with the break radius, but is measured for a larger sample
of galaxies, we also plot in Figure~\ref{fig:vrot_rbreak} (right panel) the correlation between the stellar mass and
$R_{b}$. These two parameters also show significant correlation with Pearson's coefficient of 0.705, which was also
demonstrated previously
in~\cite{Laine2016}.

\section{Conclusions}
\label{sec:conclusions}
In this work, we performed two-dimensional photometric decomposition of a sample of 375 disk galaxies viewed in the
edge-on orientation. To~perform the decomposition, we utilized $g$-, $r$- and $i$-band images from the DESI Legacy DR10
survey, which is considerably more in-depth than many older wide-field surveys---a factor crucial for this task.  Our
main results are summarized~below.

(i) The vast majority of disks in our sample have Type~II (downbending) breaks: such a feature is found in 341 galaxies
($90.9\pm 9.5\%$). The fractions of Type~II galaxies in most previous works are considerably lower ($\sim 50\%$) and do
not fall into the error margins specified here, so the difference is significant. Only one galaxy in our sample has a
Type~III (upbending) break, while the remaining 33 galaxies ($8.8 \pm 3.0\%$) are well described by a single exponential
profile (Type~I).

(ii) Such a high fraction of Type~II breaks can be attributed to greater photometric depth of the images used compared to
older wide-field surveys employed in many previous studies on the topic, and~to projection effects that make edge-on
galaxies more favorable for break~detection.

(iii) The same projection effects can lead to artificial correlations between some decomposition parameters, which
disappear when these parameters are recalculated to a face-on~orientation.

(iv) Some galaxies exhibit remarkably strong breaks with an inner-to-outer radial-scale ratio greater than 100. In~all cases, such a high ratio is due to a very large radial scale of the inner disk, which produces an almost flat light
distribution in the central~regions.

(v) Such strong breaks are separated from normal ones in some parameter planes, such as ``break strength vs. break
color'' and ``break strength vs. break surface brightness'', with almost no intermediate breaks between the two~groups.

Based on these results, we conclude that Type~II disk breaks are almost ubiquitous in disks, at~least in ones that have the same
mass and Hubble type distribution as in our sample. With~a new generation of wide-field surveys that will be 2-3
magnitudes deeper than current ones, many Type~I galaxies would reveal breaks in their outer regions that are currently
beyond the detection~limit.

The main question arising from this work concerns the nature of strong breaks. They appear to be absent in samples of
face-on or moderately inclined galaxies (or at least their fraction is considerably lower), so projection effects may
play a role, especially in combination with dust attenuation. In~this work we attempted to mitigate the impact of dust
on decomposition by masking the dust lane, but~some off-plane dust may still influence the appearance of the radial
profile. Still, the~visual inspection of galaxies with strong breaks does not suggest that such galaxies contain very
prominent dust lanes, and~the colors of their disks are not redder than those with moderate breaks. The~visual
inspection of residual maps also does not reveal that some additional components exist in such galaxies, which may
enhance the observed break~strength.

A possible way to address this problem is to perform decomposition of a sample of highly inclined---but not edge-on
---galaxies. Dependence of the strong-break fraction on the inclination in such a sample could shed light on the impact
of the projection effects on strong-break generation. This, however, would be a difficult task since the decomposition
of such galaxies faces the complication of unknown disk inclination. For edge-on galaxies, the~inclination can be
fixed to 90 degrees, while for moderately inclined ones the vertical structure of galaxies plays a less important role
and they can be adequately explained by a two-dimensional (zero-depth) exponential. Galaxies with an almost edge-on
inclination should be treated differently from face-on and perfect edge-on cases, because~one can neglect neither their
thickness nor their deviation from a perfect edge-on orientation. The~sample discussed in this work does not contain
such galaxies, and~we leave their investigation for future~work.

\funding{This research was supported by the Russian Science Foundation grant \textnumero~24--72--10084.}

\dataavailability{The decomposition results for the galaxy sample are availabe as Suplementary Materials.}

\acknowledgments{We thank the anonymous reviewers for their valuable comments and constructive suggestions, which
  significantly helped improve the quality of our~manuscript.  This research has made use of the NASA/IPAC Extragalactic
  Database (NED), which is operated by the Jet Propulsion Laboratory, California Institute of Technology, under~contract
  with the National Aeronautics and Space Administration. This research has made use of the DESI Legacy
  Survey. The~Legacy Surveys consist of three individual and complementary projects: the Dark Energy Camera Legacy
  Survey (DECaLS; Proposal ID \#2014B-0404; PIs: David Schlegel and Arjun Dey), the Beijing--Arizona Sky Survey (BASS;
  NOAO Prop. ID \#2015A-0801; PIs: Zhou Xu and Xiaohui Fan), and~the Mayall z-band Legacy Survey (MzLS; Prop. ID
  \#2016A-0453; PI: Arjun Dey). DECaLS, BASS and MzLS together include data obtained, respectively, at~the Blanco
  telescope, Cerro Tololo Inter-American Observatory, NSF’s NOIRLab; the Bok telescope, Steward Observatory, University
  of Arizona; and the Mayall telescope, Kitt Peak National Observatory, NOIRLab. Pipeline processing and analyses of the
  data were supported by NOIRLab and the Lawrence Berkeley National Laboratory (LBNL). The~Legacy Surveys project is
  honored to be permitted to conduct astronomical research on Iolkam Du’ag (Kitt Peak), a~mountain with particular
  significance to the Tohono O’odham~Nation. NOIRLab is operated by the Association of Universities for Research in
  Astronomy (AURA) under a cooperative agreement with the National Science Foundation. LBNL is managed by the Regents of
  the University of California under contract to the U.S. Department of~Energy. This project used data obtained with the
  Dark Energy Camera (DECam), which was constructed by the Dark Energy Survey (DES) collaboration. Funding for the DES
  Projects has been provided by the U.S. Department of Energy; the~ U.S. National Science Foundation; the~Ministry of
  Science and Education of Spain; the~Science and Technology Facilities Council of the United Kingdom; the~Higher
  Education Funding Council for England; the~National Center for Supercomputing Applications at the University of
  Illinois at Urbana-Champaign; the~Kavli Institute of Cosmological Physics at the University of Chicago; the Center for
  Cosmology and Astro-Particle Physics at Ohio State University; the Mitchell Institute for Fundamental Physics and
  Astronomy at Texas A\&M University; Financiadora de Estudos e Projetos; Fundacao Carlos Chagas Filho de Amparo;
  Financiadora de Estudos e Projetos; Fundacao Carlos Chagas Filho de Amparo a Pesquisa do Estado do Rio de Janeiro;
  Conselho Nacional de Desenvolvimento Cientifico e Tecnologico and the Ministerio da Ciencia; Tecnologia e Inovacao;
  the~Deutsche Forschungsgemeinschaft; and the Collaborating Institutions of the Dark Energy
  Survey.
  The~Collaborating Institutions are Argonne National Laboratory, the~University of California at Santa Cruz,
  the~University of Cambridge, Centro de Investigaciones Energeticas, Medioambientales y Tecnologicas-Madrid,
  the~University of Chicago, University College London, the~DES-Brazil Consortium, the~University of Edinburgh,
  the~Eidgenossische Technische Hochschule (ETH) Zurich, the Fermi National Accelerator Laboratory, the~ University of
  Illinois at Urbana-Champaign, the~Institut de Ciencies de l’Espai (IEEC/CSIC), the~Institut de Fisica d’Altes
  Energies, the Lawrence Berkeley National Laboratory, the~Ludwig Maximilians Universitat Munchen and the associated
  Excellence Cluster Universe, the~University of Michigan, NSF’s NOIRLab, the~University of Nottingham, Ohio State
  University, the~University of Pennsylvania, the~University of Portsmouth, the SLAC National Accelerator Laboratory,
  Stanford University, the~University of Sussex, and~Texas A\&M~University. BASS is a key project of the Telescope
  Access Program (TAP), which has been funded by the National Astronomical Observatories of China, the~Chinese Academy
  of Sciences (the Strategic Priority Research Program “The Emergence of Cosmological Structures”, Grant \#
  XDB09000000), and~the Special Fund for Astronomy from the Ministry of Finance. The~ BASS is also supported by the
  External Cooperation Program of the Chinese Academy of Sciences (Grant \# 114A11KYSB20160057), and~the Chinese
  National Natural Science Foundation (Grant \# 12120101003, \# 11433005). The Legacy Survey team makes use of data
  products from the Near-Earth Object Wide-field Infrared Survey Explorer (NEOWISE), which is a project of the Jet
  Propulsion Laboratory/California Institute of Technology. NEOWISE is funded by the National Aeronautics and
  Space~Administration.  The Legacy Surveys imaging of the DESI footprint is supported by the Director, Office of
  Science, Office of High Energy Physics of the U.S. Department of Energy under Contract No. DE-AC02-05CH1123; by~the
  National Energy Research Scientific Computing Center, a~DOE Office of Science User Facilities under the same contract;
  and by the U.S. National Science Foundation, Division of Astronomical Sciences, under Contract No. AST-0950945 to
  NOAO.}

\conflictsofinterest{The authors declare no conflicts of~interest.}

\appendixtitles{yes}
\appendixstart
\appendix
\section[\appendixname~\thesection]{Flaring Strength as a Function of Break Strength}
\label{sec:appendix_flaring}
In this section we present the derivation of Equation~(\ref{eq:flaring}), which describes the functional form of
  flaring that is required to produce an exponential break of a given strength $\alpha$. Let us consider the volume
  brightness density of a broken exponential disk beyond the break radius:
$$
\rho_{\mathrm{broken}}(x, z) = \rho (R_b) \exp\left ( -\frac{x}{h_2} \right) \exp\left ( - \frac{\left| z \right|}{h_z} \right),
$$
where $x$ is the distance from the break point, and~$\rho (R_b)$ is the volume density at the break
point. The~volume brightness density of a flaring disk in the same region would be
$$
\rho_{\mathrm{flaring}}(x, z) = \rho (R_f)\exp \left (- \frac{x}{h_1} \right)
\exp \left ( - \frac{\left|z\right|}{h_z (x)} \right) \frac{h_{z, 0}}{h_z (x)},
$$
where $R_f$ is the radius of flaring onset (by construction $R_f=R_b$), $h_{z, 0}$ is the height of the unflared disk, and~$h_z (x)$ is a
function that describes how the vertical scale depends on radius, i.e.,~the desired flaring
function.

Since we are interested in the impact of flaring on producing the light depletion in the disk
plane, we can rewrite the same equations for zero height:
$$
\rho_{\mathrm{broken}}(x) = \rho (R_b) \exp\left ( -\frac{x}{h_2} \right),
$$
and 
$$
\rho_{\mathrm{flaring}}(x) = \rho (R_f)\exp \left (- \frac{x}{h_1} \right)
\frac{h_{z, 0}}{h_z (x)}.
$$
\indent By construction, these two equations describe the same observed light distribution, so we can 
equate them:
$$
\rho (R_b) \exp\left ( -\frac{x}{h_2} \right) =  \rho (R_f)\exp \left (- \frac{x}{h_1} \right)
\frac{h_{z, 0}}{h_z (x)}.
$$
\indent Since $\rho (R_b) = \rho (R_f)$ and $h_2 = \alpha h_1$ we can rewrite this equation as
$$
\exp\left ( -\frac{x}{\alpha h_1} \right) =  \exp \left (- \frac{x}{h_1} \right)
\frac{h_{z, 0}}{h_z (x)},
$$
from which $h_z (x)$ can be inferred as
$$
h_z (x) = \exp \left( \frac{1-\alpha}{\alpha} \right) h_z (R_b).
$$

\section[\appendixname~\thesection]{Break Point Surface Brightness in Edge-on
and Face-on Orientations}
\label{ap:break_brightness}
\begin{figure}[h]
  \centering
  \includegraphics[width=0.8\linewidth]{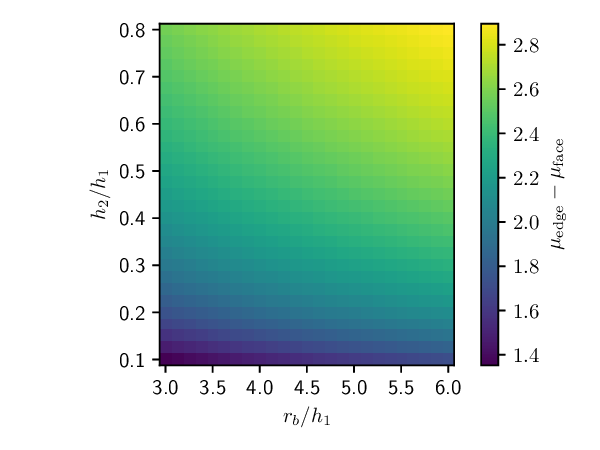}
  \caption{Difference between surface brightness in face-on and edge-on
  orientations (in magnitudes) as a function of $h_2$ and $R_b$.}
  \label{fig:face_to_edge_range}
\end{figure}

In Section~\ref{sec:breaks_fon_eon} we showed an equation that allows one to
compare the surface brightness of a disk at its break point in edge-on and
face-on orientations (equation~(\ref{eq:sb_diff_in_break})), and~demonstrated
the result of this equation for the specific combination of disk parameters
($h_2=0.25h_1$, $R_b=4h_1$, and~$h_z=0.2h_1$). Here we show the 
difference between face-on and edge-on surface brightnesses for a range of
disk parameters to demonstrate the expected strength of the effect. 
Figure~\ref{fig:face_to_edge_range} shows the results of computations for
a range of $h_2$ from 0.1 to 0.8 and a range of $R_b$ from 3.0 to 6.0 (in units
of $h_1$).

\section[\appendixname~\thesection]{The Case of PGC~731}
\label{ap:pgc_731}
\begin{figure}[h]
    \centering
    \includegraphics[width=\linewidth]{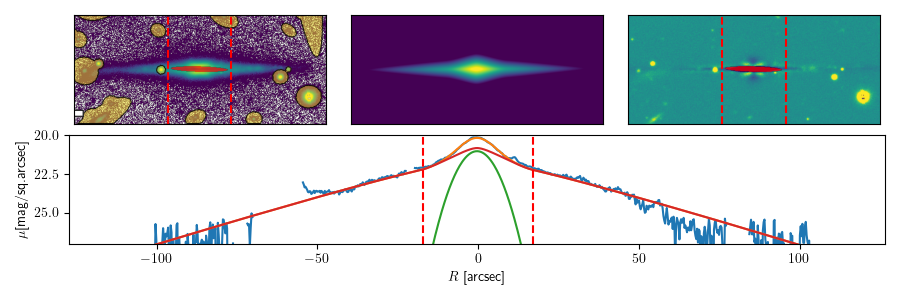}
    \caption{Decomposition results for PGC~731. Top: 2D maps as in Figure~\ref{fig:decomposition_2d}; bottom: radial slice
      as in Figure~\ref{fig:decomposition_slices}. Vertical dashed lines show break locations.}
    \label{fig:2d_decomp_1931}
  \end{figure}

In this appendix section we present the case of PGC~731, the~sole galaxy of our sample that was fitted with an
  upbending disk profile. Figure~\ref{fig:2d_decomp_1931} shows decomposition results as 2D maps (top panels) and as a
  slice along a disk plane (bottom panel). The~break location is shown by vertical dashed lines. As~can be seen from the
  figure, the~profile indeed demonstrates a break, although~it is located in the inner regions of the galaxy, at~a
  radius about two times larger than the extent of the prominent X-structure. This may indicate that the presence of a
  bar/X-structure alters the inner shape of the disk component, and~the upbending nature of this galaxy is
  questionable.

\bibliography{main}

@ARTICLE{Conselice2006,
       author = {{Conselice}, Christopher J.},
        title = "{The fundamental properties of galaxies and a new galaxy classification system}",
      journal = {\mnras},
         year = 2006,
        month = dec,
       volume = {373},
       number = {4},
        pages = {1389-1408},
          doi = {10.1111/j.1365-2966.2006.11114.x},
archivePrefix = {arXiv},
       eprint = {astro-ph/0610016},
 primaryClass = {astro-ph},
       adsurl = {https://ui.adsabs.harvard.edu/abs/2006MNRAS.373.1389C}
}

@ARTICLE{Erwin2008,
       author = {{Erwin}, Peter and {Pohlen}, Michael and {Beckman}, John E.},
        title = "{The Outer Disks of Early-Type Galaxies. I. Surface-Brightness Profiles of Barred Galaxies}",
      journal = {\aj},
         year = 2008,
        month = jan,
       volume = {135},
       number = {1},
        pages = {20-54},
          doi = {10.1088/0004-6256/135/1/20},
archivePrefix = {arXiv},
       eprint = {0709.3505},
 primaryClass = {astro-ph},
       adsurl = {https://ui.adsabs.harvard.edu/abs/2008AJ....135...20E}
}

@ARTICLE{Erwin2015,
       author = {{Erwin}, Peter},
        title = "{IMFIT: A Fast, Flexible New Program for Astronomical Image Fitting}",
      journal = {\apj},
         year = 2015,
        month = feb,
       volume = {799},
       number = {2},
          eid = {226},
        pages = {226},
          doi = {10.1088/0004-637X/799/2/226},
archivePrefix = {arXiv},
       eprint = {1408.1097},
 primaryClass = {astro-ph.IM},
       adsurl = {https://ui.adsabs.harvard.edu/abs/2015ApJ...799..226E}
}

@ARTICLE{Freeman1970,
       author = {{Freeman}, K.~C.},
        title = "{On the Disks of Spiral and S0 Galaxies}",
      journal = {\apj},
         year = 1970,
        month = jun,
       volume = {160},
        pages = {811},
          doi = {10.1086/150474},
       adsurl = {https://ui.adsabs.harvard.edu/abs/1970ApJ...160..811F}
}

@ARTICLE{Patterson1940,
       author = {{Patterson}, F. Shirley},
        title = "{The Luminosity Gradient of Messier 33}",
      journal = {Harvard College Observatory Bulletin},
         year = 1940,
        month = dec,
       volume = {914},
        pages = {9-10},
       adsurl = {https://ui.adsabs.harvard.edu/abs/1940BHarO.914....9P}
}

@ARTICLE{2022MNRAS.512.1371M,
       author = {{Marchuk}, Alexander A. and {Smirnov}, Anton A. and {Sotnikova}, Natalia Y. and {Bunakalya}, Dmitriy A. and {Savchenko}, Sergey S. and {Reshetnikov}, Vladimir P. and {Usachev}, Pavel A. and {Tikhonenko}, Iliya S. and {Zozulia}, Viktor D. and {Zakharova}, Daria A.},
        title = "{B/PS bulges in DESI Legacy edge-on galaxies - I. Sample building}",
      journal = {\mnras},
         year = 2022,
        month = may,
       volume = {512},
       number = {1},
        pages = {1371-1390},
          doi = {10.1093/mnras/stac599},
archivePrefix = {arXiv},
       eprint = {2203.01154},
 primaryClass = {astro-ph.GA},
       adsurl = {https://ui.adsabs.harvard.edu/abs/2022MNRAS.512.1371M}
}

@ARTICLE{vanderKruit1979,
       author = {{van der Kruit}, P.~C.},
        title = "{Optical surface photometry of eight spiral galaxies studied in Westerbork.}",
      journal = {\aaps},
         year = 1979,
        month = oct,
       volume = {38},
        pages = {15-38},
       adsurl = {https://ui.adsabs.harvard.edu/abs/1979A&AS...38...15V}
}

@ARTICLE{vanderKruit1981,
       author = {{van der Kruit}, P.~C. and {Searle}, L.},
        title = "{Surface photometry of edge-on spiral galaxies. I - A model for the three-dimensional distribution of light in galactic disks.}",
      journal = {\aap},
         year = 1981,
        month = feb,
       volume = {95},
        pages = {105-115},
       adsurl = {https://ui.adsabs.harvard.edu/abs/1981A&A....95..105V}
}

@INPROCEEDINGS{vanderKruit2008,
       author = {{van der Kruit}, P.~C.},
        title = "{The Stars and Gas in Outer Parts of Galaxy Disks: Extended or Truncated, Flat or Warped?}",
    booktitle = {Formation and Evolution of Galaxy Disks},
         year = 2008,
       editor = {{Funes}, J.~G. and {Corsini}, E.~M.},
       series = {Astronomical Society of the Pacific Conference Series},
       volume = {396},
        month = oct,
        pages = {173},
          doi = {10.48550/arXiv.0712.0447},
archivePrefix = {arXiv},
       eprint = {0712.0447},
 primaryClass = {astro-ph},
       adsurl = {https://ui.adsabs.harvard.edu/abs/2008ASPC..396..173V}
}

@ARTICLE{Peng2010,
       author = {{Peng}, Chien Y. and {Ho}, Luis C. and {Impey}, Chris D. and {Rix}, Hans-Walter},
        title = "{Detailed Decomposition of Galaxy Images. II. Beyond Axisymmetric Models}",
      journal = {\aj},
         year = 2010,
        month = jun,
       volume = {139},
       number = {6},
        pages = {2097-2129},
          doi = {10.1088/0004-6256/139/6/2097},
archivePrefix = {arXiv},
       eprint = {0912.0731},
 primaryClass = {astro-ph.CO},
       adsurl = {https://ui.adsabs.harvard.edu/abs/2010AJ....139.2097P}
}

@ARTICLE{Sersic1963,
       author = {{S{\'e}rsic}, J.~L.},
        title = "{Influence of the atmospheric and instrumental dispersion on the brightness distribution in a galaxy}",
      journal = {Boletin de la Asociacion Argentina de Astronomia La Plata Argentina},
         year = 1963,
        month = feb,
       volume = {6},
        pages = {41-43},
       adsurl = {https://ui.adsabs.harvard.edu/abs/1963BAAA....6...41S}
}

@ARTICLE{deVac1959,
       author = {{de Vaucouleurs}, Gerald},
        title = "{Photoelectric Photometry of Messier 33 IN the u, b, v, System.}",
      journal = {\apj},
         year = 1959,
        month = nov,
       volume = {130},
        pages = {728},
          doi = {10.1086/146764},
       adsurl = {https://ui.adsabs.harvard.edu/abs/1959ApJ...130..728D}
}

@ARTICLE{Mestel1963,
       author = {{Mestel}, L.},
        title = "{On the galactic law of rotation}",
      journal = {\mnras},
         year = 1963,
        month = jan,
       volume = {126},
        pages = {553},
          doi = {10.1093/mnras/126.6.553},
       adsurl = {https://ui.adsabs.harvard.edu/abs/1963MNRAS.126..553M}
}

@ARTICLE{Sellwood2002,
       author = {{Sellwood}, J.~A. and {Binney}, J.~J.},
        title = "{Radial mixing in galactic discs}",
      journal = {\mnras},
         year = 2002,
        month = nov,
       volume = {336},
       number = {3},
        pages = {785-796},
          doi = {10.1046/j.1365-8711.2002.05806.x},
archivePrefix = {arXiv},
       eprint = {astro-ph/0203510},
 primaryClass = {astro-ph},
       adsurl = {https://ui.adsabs.harvard.edu/abs/2002MNRAS.336..785S}
}

@ARTICLE{Roskar2008,
       author = {{Ro{\v{s}}kar}, Rok and {Debattista}, Victor P. and {Stinson}, Gregory S. and {Quinn}, Thomas R. and {Kaufmann}, Tobias and {Wadsley}, James},
        title = "{Beyond Inside-Out Growth: Formation and Evolution of Disk Outskirts}",
      journal = {\apjl},
         year = 2008,
        month = mar,
       volume = {675},
       number = {2},
        pages = {L65},
          doi = {10.1086/586734},
archivePrefix = {arXiv},
       eprint = {0710.5523},
 primaryClass = {astro-ph},
       adsurl = {https://ui.adsabs.harvard.edu/abs/2008ApJ...675L..65R}
}

@ARTICLE{Sellwood1980,
       author = {{Sellwood}, J.~A.},
        title = "{Galaxy models with live halos}",
      journal = {\aap},
         year = 1980,
        month = sep,
       volume = {89},
       number = {3},
        pages = {296-307},
       adsurl = {https://ui.adsabs.harvard.edu/abs/1980A&A....89..296S}
}

@INCOLLECTION{Athanassoula2003,
       author = {{Athanassoula}, Lia},
        title = "{Angular Momentum Redistribution and the Evolution and Morphology of Bars}",
    booktitle = {Galaxies and Chaos},
         year = 2003,
       editor = {{Contopoulos}, G. and {Voglis}, N.},
       volume = {626},
        pages = {313-326},
          doi = {10.1007/978-3-540-45040-5_26},
       adsurl = {https://ui.adsabs.harvard.edu/abs/2003LNP...626..313A}
}

@ARTICLE{Debattista2006,
       author = {{Debattista}, Victor P. and {Mayer}, Lucio and {Carollo}, C. Marcella and {Moore}, Ben and {Wadsley}, James and {Quinn}, Thomas},
        title = "{The Secular Evolution of Disk Structural Parameters}",
      journal = {\apj},
         year = 2006,
        month = jul,
       volume = {645},
       number = {1},
        pages = {209-227},
          doi = {10.1086/504147},
archivePrefix = {arXiv},
       eprint = {astro-ph/0509310},
 primaryClass = {astro-ph},
       adsurl = {https://ui.adsabs.harvard.edu/abs/2006ApJ...645..209D}
}

@ARTICLE{Goldreich1965,
       author = {{Goldreich}, P. and {Lynden-Bell}, D.},
        title = "{I. Gravitational stability of uniformly rotating disks}",
      journal = {\mnras},
         year = 1965,
        month = jan,
       volume = {130},
        pages = {97},
          doi = {10.1093/mnras/130.2.97},
       adsurl = {https://ui.adsabs.harvard.edu/abs/1965MNRAS.130...97G}
}

@ARTICLE{Kennicutt1989,
       author = {{Kennicutt}, Jr., Robert C.},
        title = "{The Star Formation Law in Galactic Disks}",
      journal = {\apj},
         year = 1989,
        month = sep,
       volume = {344},
        pages = {685},
          doi = {10.1086/167834},
       adsurl = {https://ui.adsabs.harvard.edu/abs/1989ApJ...344..685K}
}

@ARTICLE{deGrijs1997,
       author = {{de Grijs}, R. and {Peletier}, R.~F.},
        title = "{The shape of galaxy disks: how the scale height increases with galactocentric distance.}",
      journal = {\aap},
         year = 1997,
        month = apr,
       volume = {320},
        pages = {L21-L24},
          doi = {10.48550/arXiv.astro-ph/9702215},
archivePrefix = {arXiv},
       eprint = {astro-ph/9702215},
 primaryClass = {astro-ph},
       adsurl = {https://ui.adsabs.harvard.edu/abs/1997A&A...320L..21D}
}

@ARTICLE{Narayan2002,
       author = {{Narayan}, C.~A. and {Jog}, C.~J.},
        title = "{Origin of radially increasing stellar scaleheight in a galactic disk}",
      journal = {\aap},
         year = 2002,
        month = jul,
       volume = {390},
        pages = {L35-L38},
          doi = {10.1051/0004-6361:20020961},
archivePrefix = {arXiv},
       eprint = {astro-ph/0207453},
 primaryClass = {astro-ph},
       adsurl = {https://ui.adsabs.harvard.edu/abs/2002A&A...390L..35N}
}

@ARTICLE{Saha2009,
       author = {{Saha}, Kanak and {de Jong}, Roelof and {Holwerda}, Benne},
        title = "{The onset of warps in Spitzer observations of edge-on spiral galaxies}",
      journal = {\mnras},
         year = 2009,
        month = jun,
       volume = {396},
       number = {1},
        pages = {409-422},
          doi = {10.1111/j.1365-2966.2009.14696.x},
archivePrefix = {arXiv},
       eprint = {0902.4436},
 primaryClass = {astro-ph.GA},
       adsurl = {https://ui.adsabs.harvard.edu/abs/2009MNRAS.396..409S}
}

@ARTICLE{LopezCorredoria2002,
       author = {{L{\'o}pez-Corredoira}, M. and {Cabrera-Lavers}, A. and {Garz{\'o}n}, F. and {Hammersley}, P.~L.},
        title = "{Old stellar Galactic disc in near-plane regions according to 2MASS: Scales, cut-off, flare and warp}",
      journal = {\aap},
         year = 2002,
        month = nov,
       volume = {394},
        pages = {883-899},
          doi = {10.1051/0004-6361:20021175},
archivePrefix = {arXiv},
       eprint = {astro-ph/0208236},
 primaryClass = {astro-ph},
       adsurl = {https://ui.adsabs.harvard.edu/abs/2002A&A...394..883L}
}

@ARTICLE{Bakos2008,
       author = {{Bakos}, Judit and {Trujillo}, Ignacio and {Pohlen}, Michael},
        title = "{Color Profiles of Spiral Galaxies: Clues on Outer-Disk Formation Scenarios}",
      journal = {\apjl},
         year = 2008,
        month = aug,
       volume = {683},
       number = {2},
        pages = {L103},
          doi = {10.1086/591671},
archivePrefix = {arXiv},
       eprint = {0807.2776},
 primaryClass = {astro-ph},
       adsurl = {https://ui.adsabs.harvard.edu/abs/2008ApJ...683L.103B}
}

@ARTICLE{Camba2022,
       author = {{Chamba}, Nushkia and {Trujillo}, Ignacio and {Knapen}, Johan H.},
        title = "{The edges of galaxies: Tracing the limits of star formation}",
      journal = {\aap},
         year = 2022,
        month = nov,
       volume = {667},
          eid = {A87},
        pages = {A87},
          doi = {10.1051/0004-6361/202243612},
archivePrefix = {arXiv},
       eprint = {2209.05497},
 primaryClass = {astro-ph.GA},
       adsurl = {https://ui.adsabs.harvard.edu/abs/2022A&A...667A..87C}
}

@ARTICLE{Raji2025,
       author = {{Raji}, Samane and {Trujillo}, Ignacio and {Buitrago}, Fernando and {Golini}, Giulia and {Cejudo}, Ignacio Ruiz},
        title = "{Revisiting the structure of galactic disks with deep imaging}",
      journal = {\aap},
         year = 2025,
        month = dec,
       volume = {704},
          eid = {A335},
        pages = {A335},
          doi = {10.1051/0004-6361/202556488},
archivePrefix = {arXiv},
       eprint = {2510.24900},
 primaryClass = {astro-ph.GA},
       adsurl = {https://ui.adsabs.harvard.edu/abs/2025A&A...704A.335R}
}

@ARTICLE{Pohlen2006,
       author = {{Pohlen}, M. and {Trujillo}, I.},
        title = "{The structure of galactic disks. Studying late-type spiral galaxies using SDSS}",
      journal = {\aap},
         year = 2006,
        month = aug,
       volume = {454},
       number = {3},
        pages = {759-772},
          doi = {10.1051/0004-6361:20064883},
archivePrefix = {arXiv},
       eprint = {astro-ph/0603682},
 primaryClass = {astro-ph},
       adsurl = {https://ui.adsabs.harvard.edu/abs/2006A&A...454..759P}
}

@ARTICLE{Pohlen2007,
       author = {{Pohlen}, Michael and {Zaroubi}, Saleem and {Peletier}, Reynier F. and {Dettmar}, Ralf-J{\"u}rgen},
        title = "{On the three-dimensional structure of edge-on disc galaxies}",
      journal = {\mnras},
         year = 2007,
        month = jun,
       volume = {378},
       number = {2},
        pages = {594-616},
          doi = {10.1111/j.1365-2966.2007.11790.x},
archivePrefix = {arXiv},
       eprint = {astro-ph/0703768},
 primaryClass = {astro-ph},
       adsurl = {https://ui.adsabs.harvard.edu/abs/2007MNRAS.378..594P}
}

@ARTICLE{DeGeyter2014,
       author = {{De Geyter}, Gert and {Baes}, Maarten and {Camps}, Peter and {Fritz}, Jacopo and {De Looze}, Ilse and {Hughes}, Thomas M. and {Viaene}, S{\'e}bastien and {Gentile}, Gianfranco},
        title = "{The distribution of interstellar dust in CALIFA edge-on galaxies via oligochromatic radiative transfer fitting}",
      journal = {\mnras},
         year = 2014,
        month = jun,
       volume = {441},
       number = {1},
        pages = {869-885},
          doi = {10.1093/mnras/stu612},
archivePrefix = {arXiv},
       eprint = {1403.7527},
 primaryClass = {astro-ph.GA},
       adsurl = {https://ui.adsabs.harvard.edu/abs/2014MNRAS.441..869D}
}

@ARTICLE{Bianchi2007,
       author = {{Bianchi}, S.},
        title = "{The dust distribution in edge-on galaxies. Radiative transfer fits of V and K'-band images}",
      journal = {\aap},
         year = 2007,
        month = sep,
       volume = {471},
       number = {3},
        pages = {765-773},
          doi = {10.1051/0004-6361:20077649},
archivePrefix = {arXiv},
       eprint = {0705.1471},
 primaryClass = {astro-ph},
       adsurl = {https://ui.adsabs.harvard.edu/abs/2007A&A...471..765B}
}

@ARTICLE{MunozMateos2013,
       author = {{Mu{\~n}oz-Mateos}, Juan Carlos and {Sheth}, Kartik and {Gil de Paz}, Armando and {Meidt}, Sharon and {Athanassoula}, E. and {Bosma}, Albert and {Comer{\'o}n}, S{\'e}bastien and {Elmegreen}, Debra M. and {Elmegreen}, Bruce G. and {Erroz-Ferrer}, Santiago and {Gadotti}, Dimitri A. and {Hinz}, Joannah L. and {Ho}, Luis C. and {Holwerda}, Benne and {Jarrett}, Thomas H. and {Kim}, Taehyun and {Knapen}, Johan H. and {Laine}, Jarkko and {Laurikainen}, Eija and {Madore}, Barry F. and {Menendez-Delmestre}, Karin and {Mizusawa}, Trisha and {Regan}, Michael and {Salo}, Heikki and {Schinnerer}, Eva and {Seibert}, Mark and {Skibba}, Ramin and {Zaritsky}, Dennis},
        title = "{The Impact of Bars on Disk Breaks as Probed by S$^{4}$G Imaging}",
      journal = {\apj},
         year = 2013,
        month = jul,
       volume = {771},
       number = {1},
          eid = {59},
        pages = {59},
          doi = {10.1088/0004-637X/771/1/59},
archivePrefix = {arXiv},
       eprint = {1304.6083},
 primaryClass = {astro-ph.CO},
       adsurl = {https://ui.adsabs.harvard.edu/abs/2013ApJ...771...59M}
}

@ARTICLE{Gutierrez2011,
       author = {{Guti{\'e}rrez}, Leonel and {Erwin}, Peter and {Aladro}, Rebeca and {Beckman}, John E.},
        title = "{The Outer Disks of Early-type Galaxies. II. Surface-brightness Profiles of Unbarred Galaxies and Trends with Hubble Type}",
      journal = {\aj},
         year = 2011,
        month = nov,
       volume = {142},
       number = {5},
          eid = {145},
        pages = {145},
          doi = {10.1088/0004-6256/142/5/145},
archivePrefix = {arXiv},
       eprint = {1108.3662},
 primaryClass = {astro-ph.CO},
       adsurl = {https://ui.adsabs.harvard.edu/abs/2011AJ....142..145G}
}

@ARTICLE{Hammersley2011,
       author = {{Hammersley}, P.~L. and {L{\'o}pez-Corredoira}, M.},
        title = "{Modelling star counts in the Monoceros stream and the Galactic anti-centre}",
      journal = {\aap},
         year = 2011,
        month = mar,
       volume = {527},
          eid = {A6},
        pages = {A6},
          doi = {10.1051/0004-6361/200913598},
archivePrefix = {arXiv},
       eprint = {1011.2405},
 primaryClass = {astro-ph.GA},
       adsurl = {https://ui.adsabs.harvard.edu/abs/2011A&A...527A...6H}
}

@ARTICLE{Laine2014,
       author = {{Laine}, J. and {Laurikainen}, E. and {Salo}, H. and {Comer{\'o}n}, S. and {Buta}, R.~J. and {Zaritsky}, D. and {Athanassoula}, E. and {Bosma}, A. and {Mu{\~n}oz-Mateos}, J.-C. and {Gadotti}, D.~A. and {Hinz}, J.~L. and {Erroz-Ferrer}, S. and {Gil de Paz}, A. and {Kim}, T. and {Men{\'e}ndez-Delmestre}, K. and {Mizusawa}, T. and {Regan}, M.~W. and {Seibert}, M. and {Sheth}, K.},
        title = "{Morphology and environment of galaxies with disc breaks in the S$^{4}$G and NIRS0S}",
      journal = {\mnras},
         year = 2014,
        month = jul,
       volume = {441},
       number = {3},
        pages = {1992-2012},
          doi = {10.1093/mnras/stu628},
archivePrefix = {arXiv},
       eprint = {1404.0559},
 primaryClass = {astro-ph.GA},
       adsurl = {https://ui.adsabs.harvard.edu/abs/2014MNRAS.441.1992L}
}

@ARTICLE{LopezCorredoria2014,
       author = {{L{\'o}pez-Corredoira}, M. and {Molg{\'o}}, J.},
        title = "{Flare in the Galactic stellar outer disc detected in SDSS-SEGUE data}",
      journal = {\aap},
         year = 2014,
        month = jul,
       volume = {567},
          eid = {A106},
        pages = {A106},
          doi = {10.1051/0004-6361/201423706},
archivePrefix = {arXiv},
       eprint = {1405.7649},
 primaryClass = {astro-ph.GA},
       adsurl = {https://ui.adsabs.harvard.edu/abs/2014A&A...567A.106L}
}

@ARTICLE{Bovy2016,
       author = {{Bovy}, Jo and {Rix}, Hans-Walter and {Schlafly}, Edward F. and {Nidever}, David L. and {Holtzman}, Jon A. and {Shetrone}, Matthew and {Beers}, Timothy C.},
        title = "{The Stellar Population Structure of the Galactic Disk}",
      journal = {\apj},
         year = 2016,
        month = may,
       volume = {823},
       number = {1},
          eid = {30},
        pages = {30},
          doi = {10.3847/0004-637X/823/1/30},
archivePrefix = {arXiv},
       eprint = {1509.05796},
 primaryClass = {astro-ph.GA},
       adsurl = {https://ui.adsabs.harvard.edu/abs/2016ApJ...823...30B}
}

@ARTICLE{Laine2016,
       author = {{Laine}, Jarkko and {Laurikainen}, Eija and {Salo}, Heikki},
        title = "{Influence of galaxy stellar mass and observed wavelength on disc breaks in S$^{4}$G, NIRS0S, and SDSS data}",
      journal = {\aap},
         year = 2016,
        month = nov,
       volume = {596},
          eid = {A25},
        pages = {A25},
          doi = {10.1051/0004-6361/201628397},
archivePrefix = {arXiv},
       eprint = {1610.00610},
 primaryClass = {astro-ph.GA},
       adsurl = {https://ui.adsabs.harvard.edu/abs/2016A&A...596A..25L}
}

@ARTICLE{Mackereth2017,
       author = {{Mackereth}, J. Ted and {Bovy}, Jo and {Schiavon}, Ricardo P. and {Zasowski}, Gail and {Cunha}, Katia and {Frinchaboy}, Peter M. and {Garc{\'\i}a Perez}, Ana E. and {Hayden}, Michael R. and {Holtzman}, Jon and {Majewski}, Steven R. and {M{\'e}sz{\'a}ros}, Szabolcs and {Nidever}, David L. and {Pinsonneault}, Marc and {Shetrone}, Matthew D.},
        title = "{The age-metallicity structure of the Milky Way disc using APOGEE}",
      journal = {\mnras},
         year = 2017,
        month = nov,
       volume = {471},
       number = {3},
        pages = {3057-3078},
          doi = {10.1093/mnras/stx1774},
archivePrefix = {arXiv},
       eprint = {1706.00018},
 primaryClass = {astro-ph.GA},
       adsurl = {https://ui.adsabs.harvard.edu/abs/2017MNRAS.471.3057M}
}

@ARTICLE{MendezAbreu2017,
       author = {{M{\'e}ndez-Abreu}, J. and {Ruiz-Lara}, T. and {S{\'a}nchez-Menguiano}, L. and {de Lorenzo-C{\'a}ceres}, A. and {Costantin}, L. and {Catal{\'a}n-Torrecilla}, C. and {Florido}, E. and {Aguerri}, J.~A.~L. and {Bland-Hawthorn}, J. and {Corsini}, E.~M. and {Dettmar}, R.~J. and {Galbany}, L. and {Garc{\'\i}a-Benito}, R. and {Marino}, R.~A. and {M{\'a}rquez}, I. and {Ortega-Minakata}, R.~A. and {Papaderos}, P. and {S{\'a}nchez}, S.~F. and {S{\'a}nchez-Blazquez}, P. and {Spekkens}, K. and {van de Ven}, G. and {Wild}, V. and {Ziegler}, B.},
        title = "{Two-dimensional multi-component photometric decomposition of CALIFA galaxies}",
      journal = {\aap},
         year = 2017,
        month = feb,
       volume = {598},
          eid = {A32},
        pages = {A32},
          doi = {10.1051/0004-6361/201629525},
archivePrefix = {arXiv},
       eprint = {1610.05324},
 primaryClass = {astro-ph.GA},
       adsurl = {https://ui.adsabs.harvard.edu/abs/2017A&A...598A..32M}
}

@ARTICLE{Tang2020,
       author = {{Tang}, Yimeng and {Chen}, Qianhui and {Zhang}, Hong-Xin and {Lin}, Zesen and {Chen}, Guangwen and {Gao}, Yulong and {Liang}, Zhixiong and {Liu}, Haiyang and {Kong}, Xu},
        title = "{New Constraints on the Origin of Surface Brightness Profile Breaks of Disk Galaxies from MaNGA}",
      journal = {\apj},
         year = 2020,
        month = jul,
       volume = {897},
       number = {1},
          eid = {79},
        pages = {79},
          doi = {10.3847/1538-4357/ab98fd},
archivePrefix = {arXiv},
       eprint = {2006.01356},
 primaryClass = {astro-ph.GA},
       adsurl = {https://ui.adsabs.harvard.edu/abs/2020ApJ...897...79T}
}

@ARTICLE{Erwin2012,
       author = {{Erwin}, Peter and {Guti{\'e}rrez}, Leonel and {Beckman}, John E.},
        title = "{A Strong Dichotomy in S0 Disk Profiles between the Virgo Cluster and the Field}",
      journal = {\apjl},
         year = 2012,
        month = jan,
       volume = {744},
       number = {1},
          eid = {L11},
        pages = {L11},
          doi = {10.1088/2041-8205/744/1/L11},
archivePrefix = {arXiv},
       eprint = {1111.5027},
 primaryClass = {astro-ph.CO},
       adsurl = {https://ui.adsabs.harvard.edu/abs/2012ApJ...744L..11E}
}

@ARTICLE{Pranger2017,
       author = {{Pranger}, Florian and {Trujillo}, Ignacio and {Kelvin}, Lee S. and {Cebri{\'a}n}, Mar{\'\i}a},
        title = "{The effect of environment on the structure of disc galaxies}",
      journal = {\mnras},
         year = 2017,
        month = may,
       volume = {467},
       number = {2},
        pages = {2127-2144},
          doi = {10.1093/mnras/stx199},
archivePrefix = {arXiv},
       eprint = {1605.08845},
 primaryClass = {astro-ph.GA},
       adsurl = {https://ui.adsabs.harvard.edu/abs/2017MNRAS.467.2127P}
}

@ARTICLE{Mondelin2025,
       author = {{Mondelin}, M. and {Bournaud}, F. and {Cuillandre}, J.-C. and {Codis}, S. and {Stone}, C. and
                  {Bolzonella}, M. and {Sorce}, J.~G. and {Kluge}, M. and {Hatch}, N.~A. and {Marleau}, F.~R. and
                  {Schirmer}, M. and {Bouy}, H. and {Buitrago}, F. and {Tortora}, C. and {Quilley}, L. and {George},
                  K. and {Baes}, M. and {Saifollahi}, T. and {Sanchez-Alarcon}, P.~M. and {Knapen}, J.~H. and {Aghanim},
                  N. and {Amara}, A. and {Andreon}, S. and {Baccigalupi}, C. and {Balestra}, A. and {Bardelli}, S. and
                  {Battaglia}, P. and {Biviano}, A. and {Branchini}, E. and {Brescia}, M. and {Brinchmann}, J. and
                  {Capobianco}, V. and {Carbone}, C. and {Castellano}, M. and {Castignani}, G. and {Cavuoti}, S. and
                  {Cimatti}, A. and {Congedo}, G. and {Conselice}, C.~J. and {Conversi}, L. and {Copin}, Y. and
                  {Courbin}, F. and {Courtois}, H.~M. and {Cropper}, M. and {De Lucia}, G. and {Dupac}, X. and
                  {Fabricius}, M. and {Farina}, M. and {Faustini}, F. and {Ferriol}, S. and {Fotopoulou}, S. and
                  {Gillis}, B. and {Giocoli}, C. and {Grupp}, F. and {Haugan}, S.~V.~H. and {Holmes}, W. and {Hormuth},
                  F. and {Hornstrup}, A. and {Jahnke}, K. and {Jhabvala}, M. and {Keih{\"a}nen}, E. and {Kermiche},
                  S. and {Kilbinger}, M. and {Kubik}, B. and {K{\"u}mmel}, M. and {Kurki-Suonio}, H. and {Le Brun},
                  A.~M.~C. and {Ligori}, S. and {Lilje}, P.~B. and {Lindholm}, V. and {Lloro}, I. and {Maino}, D. and
                  {Maiorano}, E. and {Mansutti}, O. and {Marcin}, S. and {Marggraf}, O. and {Martinelli}, M. and
                  {Medinaceli}, E. and {Mellier}, Y. and {Merlin}, E. and {Meylan}, G. and {Moscardini}, L. and {Niemi},
                  S.-M. and {Padilla}, C. and {Pasian}, F. and {Pedersen}, K. and {Percival}, W.~J. and {Pettorino},
                  V. and {Pires}, S. and {Poncet}, M. and {Popa}, L.~A. and {Pozzetti}, L. and {Renzi}, A. and {Riccio},
                  G. and {Romelli}, E. and {Saglia}, R. and {Schneider}, P. and {Secroun}, A. and {Serrano}, S. and
                  {Sirignano}, C. and {Steinwagner}, J. and {Tereno}, I. and {Toledo-Moreo}, R. and {Torradeflot},
                  F. and {Tutusaus}, I. and {Valenziano}, L. and {Vassallo}, T. and {Wang}, Y. and {Weller}, J. and
                  {Zerbi}, F.~M. and {Zucca}, E. and {Burigana}, C. and {Scottez}, V.},
        title = "{Euclid: Early Release Observations {\textendash} The surface brightness and colour profiles of the far outskirts of galaxies in the Perseus cluster}",
      journal = {\aap},
         year = 2025,
        month = jul,
       volume = {699},
          eid = {A214},
        pages = {A214},
          doi = {10.1051/0004-6361/202554838},
archivePrefix = {arXiv},
       eprint = {2506.02745},
 primaryClass = {astro-ph.GA},
       adsurl = {https://ui.adsabs.harvard.edu/abs/2025A&A...699A.214M}
}

@ARTICLE{Pfeffer2022,
       author = {{Pfeffer}, Joel L. and {Bekki}, Kenji and {Forbes}, Duncan A. and {Couch}, Warrick J. and {Koribalski}, B{\"a}rbel S.},
        title = "{Using the EAGLE simulations to elucidate the origin of disc surface brightness profile breaks as a function of mass and environment}",
      journal = {\mnras},
         year = 2022,
        month = jan,
       volume = {509},
       number = {1},
        pages = {261-271},
          doi = {10.1093/mnras/stab2934},
archivePrefix = {arXiv},
       eprint = {2110.03856},
 primaryClass = {astro-ph.GA},
       adsurl = {https://ui.adsabs.harvard.edu/abs/2022MNRAS.509..261P}
}

@ARTICLE{Head2015,
       author = {{Head}, Jacob T.~C.~G. and {Lucey}, John R. and {Hudson}, Michael J.},
        title = "{Beyond S{\'e}rsic + exponential disc morphologies in the Coma Cluster}",
      journal = {\mnras},
         year = 2015,
        month = nov,
       volume = {453},
       number = {4},
        pages = {3729-3753},
          doi = {10.1093/mnras/stv1662},
archivePrefix = {arXiv},
       eprint = {1507.07930},
 primaryClass = {astro-ph.GA},
       adsurl = {https://ui.adsabs.harvard.edu/abs/2015MNRAS.453.3729H}
}

@ARTICLE{MartinNavarro2012,
       author = {{Mart{\'\i}n-Navarro}, Ignacio and {Bakos}, Judit and {Trujillo}, Ignacio and {Knapen}, Johan H. and {Athanassoula}, E. and {Bosma}, Albert and {Comer{\'o}n}, S{\'e}bastien and {Elmegreen}, Bruce G. and {Erroz-Ferrer}, Santiago and {Gadotti}, Dimitri A. and {Gil de Paz}, Armando and {Hinz}, Joannah L. and {Ho}, Luis C. and {Holwerda}, Benne W. and {Kim}, Taehyun and {Laine}, Jarkko and {Laurikainen}, Eija and {Men{\'e}ndez-Delmestre}, Kar{\'\i}n. and {Mizusawa}, Trisha and {Mu{\~n}oz-Mateos}, Juan-Carlos and {Regan}, Michael W. and {Salo}, Heikki and {Seibert}, Mark and {Sheth}, Kartik},
        title = "{A unified picture of breaks and truncations in spiral galaxies from SDSS and S$^{4}$G imaging}",
      journal = {\mnras},
         year = 2012,
        month = dec,
       volume = {427},
       number = {2},
        pages = {1102-1134},
          doi = {10.1111/j.1365-2966.2012.21929.x},
archivePrefix = {arXiv},
       eprint = {1208.2893},
 primaryClass = {astro-ph.CO},
       adsurl = {https://ui.adsabs.harvard.edu/abs/2012MNRAS.427.1102M}
}

@ARTICLE{Dey2019,
       author = {{Dey}, Arjun and {Schlegel}, David J. and {Lang}, Dustin and {Blum}, Robert and {Burleigh}, Kaylan and {Fan}, Xiaohui and {Findlay}, Joseph R. and {Finkbeiner}, Doug and {Herrera}, David and {Juneau}, St{\'e}phanie and {Landriau}, Martin and {Levi}, Michael and {McGreer}, Ian and {Meisner}, Aaron and {Myers}, Adam D. and {Moustakas}, John and {Nugent}, Peter and {Patej}, Anna and {Schlafly}, Edward F. and {Walker}, Alistair R. and {Valdes}, Francisco and {Weaver}, Benjamin A. and {Y{\`e}che}, Christophe and {Zou}, Hu and {Zhou}, Xu and {Abareshi}, Behzad and {Abbott}, T.~M.~C. and {Abolfathi}, Bela and {Aguilera}, C. and {Alam}, Shadab and {Allen}, Lori and {Alvarez}, A. and {Annis}, James and {Ansarinejad}, Behzad and {Aubert}, Marie and {Beechert}, Jacqueline and {Bell}, Eric F. and {BenZvi}, Segev Y. and {Beutler}, Florian and {Bielby}, Richard M. and {Bolton}, Adam S. and {Brice{\~n}o}, C{\'e}sar and {Buckley-Geer}, Elizabeth J. and {Butler}, Karen and {Calamida}, Annalisa and {Carlberg}, Raymond G. and {Carter}, Paul and {Casas}, Ricard and {Castander}, Francisco J. and {Choi}, Yumi and {Comparat}, Johan and {Cukanovaite}, Elena and {Delubac}, Timoth{\'e}e and {DeVries}, Kaitlin and {Dey}, Sharmila and {Dhungana}, Govinda and {Dickinson}, Mark and {Ding}, Zhejie and {Donaldson}, John B. and {Duan}, Yutong and {Duckworth}, Christopher J. and {Eftekharzadeh}, Sarah and {Eisenstein}, Daniel J. and {Etourneau}, Thomas and {Fagrelius}, Parker A. and {Farihi}, Jay and {Fitzpatrick}, Mike and {Font-Ribera}, Andreu and {Fulmer}, Leah and {G{\"a}nsicke}, Boris T. and {Gaztanaga}, Enrique and {George}, Koshy and {Gerdes}, David W. and {Gontcho}, Satya Gontcho A. and {Gorgoni}, Claudio and {Green}, Gregory and {Guy}, Julien and {Harmer}, Diane and {Hernandez}, M. and {Honscheid}, Klaus and {Huang}, Lijuan Wendy and {James}, David J. and {Jannuzi}, Buell T. and {Jiang}, Linhua and {Joyce}, Richard and {Karcher}, Armin and {Karkar}, Sonia and {Kehoe}, Robert and {Kneib}, Jean-Paul and {Kueter-Young}, Andrea and {Lan}, Ting-Wen and {Lauer}, Tod R. and {Le Guillou}, Laurent and {Le Van Suu}, Auguste and {Lee}, Jae Hyeon and {Lesser}, Michael and {Perreault Levasseur}, Laurence and {Li}, Ting S. and {Mann}, Justin L. and {Marshall}, Robert and {Mart{\'\i}nez-V{\'a}zquez}, C.~E. and {Martini}, Paul and {du Mas des Bourboux}, H{\'e}lion and {McManus}, Sean and {Meier}, Tobias Gabriel and {M{\'e}nard}, Brice and {Metcalfe}, Nigel and {Mu{\~n}oz-Guti{\'e}rrez}, Andrea and {Najita}, Joan and {Napier}, Kevin and {Narayan}, Gautham and {Newman}, Jeffrey A. and {Nie}, Jundan and {Nord}, Brian and {Norman}, Dara J. and {Olsen}, Knut A.~G. and {Paat}, Anthony and {Palanque-Delabrouille}, Nathalie and {Peng}, Xiyan and {Poppett}, Claire L. and {Poremba}, Megan R. and {Prakash}, Abhishek and {Rabinowitz}, David and {Raichoor}, Anand and {Rezaie}, Mehdi and {Robertson}, A.~N. and {Roe}, Natalie A. and {Ross}, Ashley J. and {Ross}, Nicholas P. and {Rudnick}, Gregory and {Safonova}, Sasha and {Saha}, Abhijit and {S{\'a}nchez}, F. Javier and {Savary}, Elodie and {Schweiker}, Heidi and {Scott}, Adam and {Seo}, Hee-Jong and {Shan}, Huanyuan and {Silva}, David R. and {Slepian}, Zachary and {Soto}, Christian and {Sprayberry}, David and {Staten}, Ryan and {Stillman}, Coley M. and {Stupak}, Robert J. and {Summers}, David L. and {Sien Tie}, Suk and {Tirado}, H. and {Vargas-Maga{\~n}a}, Mariana and {Vivas}, A. Katherina and {Wechsler}, Risa H. and {Williams}, Doug and {Yang}, Jinyi and {Yang}, Qian and {Yapici}, Tolga and {Zaritsky}, Dennis and {Zenteno}, A. and {Zhang}, Kai and {Zhang}, Tianmeng and {Zhou}, Rongpu and {Zhou}, Zhimin},
        title = "{Overview of the DESI Legacy Imaging Surveys}",
      journal = {\aj},
         year = 2019,
        month = may,
       volume = {157},
       number = {5},
          eid = {168},
        pages = {168},
          doi = {10.3847/1538-3881/ab089d},
archivePrefix = {arXiv},
       eprint = {1804.08657},
 primaryClass = {astro-ph.IM},
       adsurl = {https://ui.adsabs.harvard.edu/abs/2019AJ....157..168D}
}

@ARTICLE{Bertin1996,
   author = {{Bertin}, E. and {Arnouts}, S.},
    title = "{SExtractor: Software for source extraction.}",
  journal = {\aaps},
     year = 1996,
    month = jun,
   volume = 117,
    pages = {393-404},
      doi = {10.1051/aas:1996164},
   adsurl = {https://ui.adsabs.harvard.edu/abs/1996A%26AS..117..393B}
}

@INPROCEEDINGS{Joye2003,
       author = {{Joye}, W.~A. and {Mandel}, E.},
        title = "{New Features of SAOImage DS9}",
    booktitle = {Astronomical Data Analysis Software and Systems XII},
         year = 2003,
       editor = {{Payne}, H.~E. and {Jedrzejewski}, R.~I. and {Hook}, R.~N.},
       series = {Astronomical Society of the Pacific Conference Series},
       volume = {295},
        month = jan,
        pages = {489},
       adsurl = {https://ui.adsabs.harvard.edu/abs/2003ASPC..295..489J}
}

@ARTICLE{Kasparova2020,
       author = {{Kasparova}, Anastasia V. and {Katkov}, Ivan Yu and {Chilingarian}, Igor V.},
        title = "{An excessively massive thick disc of the enormous edge-on lenticular galaxy NGC 7572}",
      journal = {\mnras},
         year = 2020,
        month = apr,
       volume = {493},
       number = {4},
        pages = {5464-5478},
          doi = {10.1093/mnras/staa611},
archivePrefix = {arXiv},
       eprint = {1912.04887},
 primaryClass = {astro-ph.GA},
       adsurl = {https://ui.adsabs.harvard.edu/abs/2020MNRAS.493.5464K}
}

@ARTICLE{Gadotti2010,
       author = {{Gadotti}, Dimitri A. and {Baes}, Maarten and {Falony}, Sarah},
        title = "{Radiative transfer in disc galaxies - IV. The effects of dust attenuation on bulge and disc structural parameters}",
      journal = {\mnras},
         year = 2010,
        month = apr,
       volume = {403},
       number = {4},
        pages = {2053-2062},
          doi = {10.1111/j.1365-2966.2010.16243.x},
archivePrefix = {arXiv},
       eprint = {1001.2303},
 primaryClass = {astro-ph.CO},
       adsurl = {https://ui.adsabs.harvard.edu/abs/2010MNRAS.403.2053G}
}

@ARTICLE{Pastrav2013a,
       author = {{Pastrav}, B.~A. and {Popescu}, C.~C. and {Tuffs}, R.~J. and {Sansom}, A.~E.},
        title = "{The effects of dust on the derived photometric parameters of disks and bulges in spiral galaxies}",
      journal = {\aap},
         year = 2013,
        month = may,
       volume = {553},
          eid = {A80},
        pages = {A80},
          doi = {10.1051/0004-6361/201220962},
archivePrefix = {arXiv},
       eprint = {1301.5602},
 primaryClass = {astro-ph.CO},
       adsurl = {https://ui.adsabs.harvard.edu/abs/2013A&A...553A..80P}
}

@ARTICLE{Pastrav2013b,
       author = {{Pastrav}, Bogdan A. and {Popescu}, Cristina C. and {Tuffs}, Richard J. and {Sansom}, Anne E.},
        title = "{The effects of dust on the photometric parameters of decomposed disks and bulges}",
      journal = {\aap},
         year = 2013,
        month = sep,
       volume = {557},
          eid = {A137},
        pages = {A137},
          doi = {10.1051/0004-6361/201322086},
archivePrefix = {arXiv},
       eprint = {1308.0458},
 primaryClass = {astro-ph.CO},
       adsurl = {https://ui.adsabs.harvard.edu/abs/2013A&A...557A.137P}
}

@ARTICLE{Savchenko2023,
       author = {{Savchenko}, Sergey S. and {Poliakov}, Denis M. and {Mosenkov}, Aleksandr V. and {Smirnov}, Anton A. and {Marchuk}, Alexander A. and {Il'in}, Vladimir B. and {Gontcharov}, George A. and {Seguine}, Jonah and {Baes}, Maarten},
        title = "{The problem of dust attenuation in photometric decomposition of edge-on galaxies and possible solutions}",
      journal = {\mnras},
         year = 2023,
        month = sep,
       volume = {524},
       number = {3},
        pages = {4729-4745},
          doi = {10.1093/mnras/stad2189},
archivePrefix = {arXiv},
       eprint = {2309.06257},
 primaryClass = {astro-ph.GA},
       adsurl = {https://ui.adsabs.harvard.edu/abs/2023MNRAS.524.4729S}
}

@ARTICLE{Makarov2022,
       author = {{Makarov}, D. and {Savchenko}, S. and {Mosenkov}, A. and {Bizyaev}, D. and {Reshetnikov}, V. and {Antipova}, A. and {Tikhonenko}, I. and {Usachev}, P. and {Borisov}, S. and {Makarova}, L. and {Kautsch}, S. and {Marchuk}, A. and {Rubtsov}, E.},
        title = "{The edge-on Galaxies in the Pan-STARRS survey (EGIPS)}",
      journal = {\mnras},
         year = 2022,
        month = apr,
       volume = {511},
       number = {2},
        pages = {3063-3075},
          doi = {10.1093/mnras/stac227},
archivePrefix = {arXiv},
       eprint = {2201.08888},
 primaryClass = {astro-ph.GA},
       adsurl = {https://ui.adsabs.harvard.edu/abs/2022MNRAS.511.3063M}
}

@ARTICLE{Karachentsev1999,
       author = {{Karachentsev}, I.~D. and {Karachentseva}, V.~E. and {Kudrya}, Yu. N. and {Sharina}, M.~E. and {Parnovskij}, S.~L.},
        title = "{The revised Flat Galaxy Catalogue.}",
      journal = {Bulletin of the Special Astrophysics Observatory},
         year = 1999,
        month = jan,
       volume = {47},
        pages = {5-185},
          doi = {10.48550/arXiv.astro-ph/0305566},
archivePrefix = {arXiv},
       eprint = {astro-ph/0305566},
 primaryClass = {astro-ph},
       adsurl = {https://ui.adsabs.harvard.edu/abs/1999BSAO...47....5K}
}

@ARTICLE{Bizyaev2004,
       author = {{Bizyaev}, D. and {Kajsin}, S.},
        title = "{The Stellar Disk Thickness of Low Surface Brightness Galaxies}",
      journal = {\apj},
         year = 2004,
        month = oct,
       volume = {613},
       number = {2},
        pages = {886-897},
          doi = {10.1086/423229},
archivePrefix = {arXiv},
       eprint = {astro-ph/0406498},
 primaryClass = {astro-ph},
       adsurl = {https://ui.adsabs.harvard.edu/abs/2004ApJ...613..886B}
}

@ARTICLE{Kregel2002,
       author = {{Kregel}, Michiel and {van der Kruit}, Pieter C. and {de Grijs}, Richard},
        title = "{Flattening and truncation of stellar discs in edge-on spiral galaxies}",
      journal = {\mnras},
         year = 2002,
        month = aug,
       volume = {334},
       number = {3},
        pages = {646-668},
          doi = {10.1046/j.1365-8711.2002.05556.x},
archivePrefix = {arXiv},
       eprint = {astro-ph/0204154},
 primaryClass = {astro-ph},
       adsurl = {https://ui.adsabs.harvard.edu/abs/2002MNRAS.334..646K}
}

@ARTICLE{deGrijs1998,
       author = {{de Grijs}, R.},
        title = "{The global structure of galactic discs}",
      journal = {\mnras},
         year = 1998,
        month = sep,
       volume = {299},
       number = {2},
        pages = {595-610},
          doi = {10.1046/j.1365-8711.1998.01896.x},
archivePrefix = {arXiv},
       eprint = {astro-ph/9804337},
 primaryClass = {astro-ph},
       adsurl = {https://ui.adsabs.harvard.edu/abs/1998MNRAS.299..595D}
}

@ARTICLE{Borlaff2016,
       author = {{Borlaff}, Alejandro and {Eliche-Moral}, M. Carmen and {Beckman}, John and {Font}, Joan},
        title = "{Type-II surface brightness profiles in edge-on galaxies produced by flares}",
      journal = {\aap},
         year = 2016,
        month = jun,
       volume = {591},
          eid = {L7},
        pages = {L7},
          doi = {10.1051/0004-6361/201628868},
archivePrefix = {arXiv},
       eprint = {1606.00448},
 primaryClass = {astro-ph.GA},
       adsurl = {https://ui.adsabs.harvard.edu/abs/2016A&A...591L...7B}
}

@ARTICLE{Comeron2012,
       author = {{Comer{\'o}n}, S{\'e}bastien and {Elmegreen}, Bruce G. and {Salo}, Heikki and {Laurikainen}, Eija and {Athanassoula}, E. and {Bosma}, Albert and {Knapen}, Johan H. and {Gadotti}, Dimitri A. and {Sheth}, Kartik and {Hinz}, Joannah L. and {Regan}, Michael W. and {Gil de Paz}, Armando and {Mu{\~n}oz-Mateos}, Juan Carlos and {Men{\'e}ndez-Delmestre}, Kar{\'\i}n and {Seibert}, Mark and {Kim}, Taehyun and {Mizusawa}, Trisha and {Laine}, Jarkko and {Ho}, Luis C. and {Holwerda}, Benne},
        title = "{Breaks in Thin and Thick Disks of Edge-on Galaxies Imaged in the Spitzer Survey Stellar Structure in Galaxies (S$^{4}$G)}",
      journal = {\apj},
         year = 2012,
        month = nov,
       volume = {759},
       number = {2},
          eid = {98},
        pages = {98},
          doi = {10.1088/0004-637X/759/2/98},
archivePrefix = {arXiv},
       eprint = {1209.1513},
 primaryClass = {astro-ph.CO},
       adsurl = {https://ui.adsabs.harvard.edu/abs/2012ApJ...759...98C}
}

@article{Comeron2018,
  title = {The Reports of Thick Discs' Deaths Are Greatly Exaggerated. {{Thick}} Discs Are {{NOT}} Artefacts Caused by Diffuse Scattered Light},
  author = {Comer{\'o}n, S. and Salo, H. and Knapen, J. H.},
  year = 2018,
  month = feb,
  journal = {Astronomy and Astrophysics},
  volume = {610},
  pages = {A5},
  publisher = {EDP},
  issn = {0004-6361},
  doi = {10.1051/0004-6361/201731415}
}

@ARTICLE{Smirnov2026,
       author = {{Smirnov}, Anton and {Marchuk}, Alexander and {Zozulia}, Viktor and {Sotnikova}, Natalia and {Savchenko}, Sergey},
        title = "{Boxy/Peanut Bulges: Comparative Analysis of EGIPS Galaxies and TNG50 Models}",
      journal = {Galaxies},
         year = 2026,
        month = jan,
       volume = {14},
       number = {1},
          eid = {4},
        pages = {4},
          doi = {10.3390/galaxies14010004},
archivePrefix = {arXiv},
       eprint = {2601.13893},
 primaryClass = {astro-ph.GA},
       adsurl = {https://ui.adsabs.harvard.edu/abs/2026Galax..14....4S}
}

@ARTICLE{Combes1981,
       author = {{Combes}, F. and {Sanders}, R.~H.},
        title = "{Formation and properties of persisting stellar bars.}",
      journal = {\aap},
         year = 1981,
        month = mar,
       volume = {96},
        pages = {164-173},
       adsurl = {https://ui.adsabs.harvard.edu/abs/1981A&A....96..164C}
}

@ARTICLE{Raha1991,
       author = {{Raha}, N. and {Sellwood}, J.~A. and {James}, R.~A. and {Kahn}, F.~D.},
        title = "{A dynamical instability of bars in disk galaxies}",
      journal = {\nat},
         year = 1991,
        month = aug,
       volume = {352},
       number = {6334},
        pages = {411-412},
          doi = {10.1038/352411a0},
       adsurl = {https://ui.adsabs.harvard.edu/abs/1991Natur.352..411R}
}

@ARTICLE{Smirnov2020,
       author = {{Smirnov}, Anton A. and {Savchenko}, Sergey S.},
        title = "{New X-shaped bulge photometric model as a tool for measuring B/PS bulges and their X-structures in photometric studies}",
      journal = {\mnras},
         year = 2020,
        month = nov,
       volume = {499},
       number = {1},
        pages = {462-481},
          doi = {10.1093/mnras/staa2892},
archivePrefix = {arXiv},
       eprint = {2007.12121},
 primaryClass = {astro-ph.GA},
       adsurl = {https://ui.adsabs.harvard.edu/abs/2020MNRAS.499..462S}
}

@ARTICLE{Savchenko2017,
       author = {{Savchenko}, S.~S. and {Sotnikova}, N. Ya. and {Mosenkov}, A.~V. and {Reshetnikov}, V.~P. and {Bizyaev}, D.~V.},
        title = "{Measuring the X-shaped structures in edge-on galaxies}",
      journal = {\mnras},
         year = 2017,
        month = nov,
       volume = {471},
       number = {3},
        pages = {3261-3272},
          doi = {10.1093/mnras/stx1802},
archivePrefix = {arXiv},
       eprint = {1707.04700},
 primaryClass = {astro-ph.GA},
       adsurl = {https://ui.adsabs.harvard.edu/abs/2017MNRAS.471.3261S}
}

@ARTICLE{Chen2026,
       author = {{Chen}, Liufei and {Du}, Min and {Lu}, Shuai and {Li}, Jing and {Ho}, Luis C.},
        title = "{Down-bending Breaks in Galactic Disks Are an Intrinsic Byproduct of Inside-out Growth}",
      journal = {arXiv e-prints},
         year = 2026,
        month = jan,
          eid = {arXiv:2602.00626},
        pages = {arXiv:2602.00626},
          doi = {10.48550/arXiv.2602.00626},
archivePrefix = {arXiv},
       eprint = {2602.00626},
 primaryClass = {astro-ph.GA},
       adsurl = {https://ui.adsabs.harvard.edu/abs/2026arXiv260200626C}
}

@ARTICLE{Nelson2019,
       author = {{Nelson}, Dylan and {Springel}, Volker and {Pillepich}, Annalisa and {Rodriguez-Gomez}, Vicente and {Torrey}, Paul and {Genel}, Shy and {Vogelsberger}, Mark and {Pakmor}, Ruediger and {Marinacci}, Federico and {Weinberger}, Rainer and {Kelley}, Luke and {Lovell}, Mark and {Diemer}, Benedikt and {Hernquist}, Lars},
        title = "{The IllustrisTNG simulations: public data release}",
      journal = {Computational Astrophysics and Cosmology},
         year = 2019,
        month = may,
       volume = {6},
       number = {1},
          eid = {2},
        pages = {2},
          doi = {10.1186/s40668-019-0028-x},
archivePrefix = {arXiv},
       eprint = {1812.05609},
 primaryClass = {astro-ph.GA},
       adsurl = {https://ui.adsabs.harvard.edu/abs/2019ComAC...6....2N}
}

@ARTICLE{Schlafly2011,
       author = {{Schlafly}, Edward F. and {Finkbeiner}, Douglas P.},
        title = "{Measuring Reddening with Sloan Digital Sky Survey Stellar Spectra and Recalibrating SFD}",
      journal = {\apj},
         year = 2011,
        month = aug,
       volume = {737},
       number = {2},
          eid = {103},
        pages = {103},
          doi = {10.1088/0004-637X/737/2/103},
archivePrefix = {arXiv},
       eprint = {1012.4804},
 primaryClass = {astro-ph.GA},
       adsurl = {https://ui.adsabs.harvard.edu/abs/2011ApJ...737..103S}
}

@ARTICLE{Sheth2010,
       author = {{Sheth}, Kartik and {Regan}, Michael and {Hinz}, Joannah L. and {Gil de Paz}, Armando and {Men{\'e}ndez-Delmestre}, Kar{\'\i}n and {Mu{\~n}oz-Mateos}, Juan-Carlos and {Seibert}, Mark and {Kim}, Taehyun and {Laurikainen}, Eija and {Salo}, Heikki and {Gadotti}, Dimitri A. and {Laine}, Jarkko and {Mizusawa}, Trisha and {Armus}, Lee and {Athanassoula}, E. and {Bosma}, Albert and {Buta}, Ronald J. and {Capak}, Peter and {Jarrett}, Thomas H. and {Elmegreen}, Debra M. and {Elmegreen}, Bruce G. and {Knapen}, Johan H. and {Koda}, Jin and {Helou}, George and {Ho}, Luis C. and {Madore}, Barry F. and {Masters}, Karen L. and {Mobasher}, Bahram and {Ogle}, Patrick and {Peng}, Chien Y. and {Schinnerer}, Eva and {Surace}, Jason A. and {Zaritsky}, Dennis and {Comer{\'o}n}, S{\'e}bastien and {de Swardt}, Bonita and {Meidt}, Sharon E. and {Kasliwal}, Mansi and {Aravena}, Manuel},
        title = "{The Spitzer Survey of Stellar Structure in Galaxies (S4G)}",
      journal = {\pasp},
         year = 2010,
        month = dec,
       volume = {122},
       number = {898},
        pages = {1397-1414},
          doi = {10.1086/657638},
archivePrefix = {arXiv},
       eprint = {1010.1592},
 primaryClass = {astro-ph.CO},
       adsurl = {https://ui.adsabs.harvard.edu/abs/2010PASP..122.1397S}
}

@ARTICLE{Bakos2012,
       author = {{Bakos}, Judit and {Trujillo}, Ignacio},
        title = "{Deep Surface Brightness Profiles of Spiral Galaxies from SDSS Stripe82: Touching Stellar Halos}",
      journal = {arXiv e-prints},
         year = 2012,
        month = apr,
          eid = {arXiv:1204.3082},
        pages = {arXiv:1204.3082},
          doi = {10.48550/arXiv.1204.3082},
archivePrefix = {arXiv},
       eprint = {1204.3082},
 primaryClass = {astro-ph.CO},
       adsurl = {https://ui.adsabs.harvard.edu/abs/2012arXiv1204.3082B}
}

@ARTICLE{Jiang2014,
       author = {{Jiang}, Linhua and {Fan}, Xiaohui and {Bian}, Fuyan and {McGreer}, Ian D. and {Strauss}, Michael A. and {Annis}, James and {Buck}, Zo{\"e} and {Green}, Richard and {Hodge}, Jacqueline A. and {Myers}, Adam D. and {Rafiee}, Alireza and {Richards}, Gordon},
        title = "{The Sloan Digital Sky Survey Stripe 82 Imaging Data: Depth-optimized Co-adds over 300 deg$^{2}$ in Five Filters}",
      journal = {\apjs},
         year = 2014,
        month = jul,
       volume = {213},
       number = {1},
          eid = {12},
        pages = {12},
          doi = {10.1088/0067-0049/213/1/12},
archivePrefix = {arXiv},
       eprint = {1405.7382},
 primaryClass = {astro-ph.GA},
       adsurl = {https://ui.adsabs.harvard.edu/abs/2014ApJS..213...12J}
}

@ARTICLE{Maltby2012,
       author = {{Maltby}, David T. and {Gray}, Meghan E. and {Arag{\'o}n-Salamanca}, Alfonso and {Wolf}, Christian and {Bell}, Eric F. and {Jogee}, Shardha and {H{\"a}u{\ss}ler}, Boris and {Barazza}, Fabio D. and {B{\"o}hm}, Asmus and {Jahnke}, Knud},
        title = "{The environmental dependence of the structure of outer galactic discs in STAGES spiral galaxies}",
      journal = {\mnras},
         year = 2012,
        month = jan,
       volume = {419},
       number = {1},
        pages = {669-686},
          doi = {10.1111/j.1365-2966.2011.19727.x},
archivePrefix = {arXiv},
       eprint = {1108.6206},
 primaryClass = {astro-ph.CO},
       adsurl = {https://ui.adsabs.harvard.edu/abs/2012MNRAS.419..669M}
}

@ARTICLE{Erwin2005,
       author = {{Erwin}, Peter and {Beckman}, John E. and {Pohlen}, Michael},
        title = "{Antitruncation of Disks in Early-Type Barred Galaxies}",
      journal = {\apjl},
         year = 2005,
        month = jun,
       volume = {626},
       number = {2},
        pages = {L81-L84},
          doi = {10.1086/431739},
archivePrefix = {arXiv},
       eprint = {astro-ph/0505216},
 primaryClass = {astro-ph},
       adsurl = {https://ui.adsabs.harvard.edu/abs/2005ApJ...626L..81E}
}

@ARTICLE{Chambers2016,
       author = {{Chambers}, K.~C. and {Magnier}, E.~A. and {Metcalfe}, N. and {Flewelling}, H.~A. and {Huber}, M.~E. and {Waters}, C.~Z. and {Denneau}, L. and {Draper}, P.~W. and {Farrow}, D. and {Finkbeiner}, D.~P. and {Holmberg}, C. and {Koppenhoefer}, J. and {Price}, P.~A. and {Rest}, A. and {Saglia}, R.~P. and {Schlafly}, E.~F. and {Smartt}, S.~J. and {Sweeney}, W. and {Wainscoat}, R.~J. and {Burgett}, W.~S. and {Chastel}, S. and {Grav}, T. and {Heasley}, J.~N. and {Hodapp}, K.~W. and {Jedicke}, R. and {Kaiser}, N. and {Kudritzki}, R.-P. and {Luppino}, G.~A. and {Lupton}, R.~H. and {Monet}, D.~G. and {Morgan}, J.~S. and {Onaka}, P.~M. and {Shiao}, B. and {Stubbs}, C.~W. and {Tonry}, J.~L. and {White}, R. and {Ba{\~n}ados}, E. and {Bell}, E.~F. and {Bender}, R. and {Bernard}, E.~J. and {Boegner}, M. and {Boffi}, F. and {Botticella}, M.~T. and {Calamida}, A. and {Casertano}, S. and {Chen}, W.-P. and {Chen}, X. and {Cole}, S. and {Deacon}, N. and {Frenk}, C. and {Fitzsimmons}, A. and {Gezari}, S. and {Gibbs}, V. and {Goessl}, C. and {Goggia}, T. and {Gourgue}, R. and {Goldman}, B. and {Grant}, P. and {Grebel}, E.~K. and {Hambly}, N.~C. and {Hasinger}, G. and {Heavens}, A.~F. and {Heckman}, T.~M. and {Henderson}, R. and {Henning}, T. and {Holman}, M. and {Hopp}, U. and {Ip}, W.-H. and {Isani}, S. and {Jackson}, M. and {Keyes}, C.~D. and {Koekemoer}, A.~M. and {Kotak}, R. and {Le}, D. and {Liska}, D. and {Long}, K.~S. and {Lucey}, J.~R. and {Liu}, M. and {Martin}, N.~F. and {Masci}, G. and {McLean}, B. and {Mindel}, E. and {Misra}, P. and {Morganson}, E. and {Murphy}, D.~N.~A. and {Obaika}, A. and {Narayan}, G. and {Nieto-Santisteban}, M.~A. and {Norberg}, P. and {Peacock}, J.~A. and {Pier}, E.~A. and {Postman}, M. and {Primak}, N. and {Rae}, C. and {Rai}, A. and {Riess}, A. and {Riffeser}, A. and {Rix}, H.~W. and {R{\"o}ser}, S. and {Russel}, R. and {Rutz}, L. and {Schilbach}, E. and {Schultz}, A.~S.~B. and {Scolnic}, D. and {Strolger}, L. and {Szalay}, A. and {Seitz}, S. and {Small}, E. and {Smith}, K.~W. and {Soderblom}, D.~R. and {Taylor}, P. and {Thomson}, R. and {Taylor}, A.~N. and {Thakar}, A.~R. and {Thiel}, J. and {Thilker}, D. and {Unger}, D. and {Urata}, Y. and {Valenti}, J. and {Wagner}, J. and {Walder}, T. and {Walter}, F. and {Watters}, S.~P. and {Werner}, S. and {Wood-Vasey}, W.~M. and {Wyse}, R.},
        title = "{The Pan-STARRS1 Surveys}",
      journal = {arXiv e-prints},
         year = 2016,
        month = dec,
          eid = {arXiv:1612.05560},
        pages = {arXiv:1612.05560},
          doi = {10.48550/arXiv.1612.05560},
archivePrefix = {arXiv},
       eprint = {1612.05560},
 primaryClass = {astro-ph.IM},
       adsurl = {https://ui.adsabs.harvard.edu/abs/2016arXiv161205560C}
}

@ARTICLE{Bell2003,
       author = {{Bell}, Eric F. and {McIntosh}, Daniel H. and {Katz}, Neal and {Weinberg}, Martin D.},
        title = "{The Optical and Near-Infrared Properties of Galaxies. I. Luminosity and Stellar Mass Functions}",
      journal = {\apjs},
         year = 2003,
        month = dec,
       volume = {149},
       number = {2},
        pages = {289-312},
          doi = {10.1086/378847},
archivePrefix = {arXiv},
       eprint = {astro-ph/0302543},
 primaryClass = {astro-ph},
       adsurl = {https://ui.adsabs.harvard.edu/abs/2003ApJS..149..289B}
}

@Article{Makarov2014,
  author	= {{Makarov}, D. and {Prugniel}, P. and {Terekhova}, N. and
		  {Courtois}, H. and {Vauglin}, I.},
  title		= "{HyperLEDA. III. The catalogue of extragalactic
		  distances}",
  journal	= {\aap},
  year		= 2014,
  month		= oct,
  volume	= 570,
  eid		= {A13},
  pages		= {A13},
  doi		= {10.1051/0004-6361/201423496},
  adsurl	= {http://adsabs.harvard.edu/abs/2014A%26A...570A..13M}
}

@ARTICLE{Ebrova2025,
       author = {{Ebrov{\'a}}, Ivana and {B{\'\i}lek}, Michal and {Eli{\'a}{\v{s}}ek}, Ji{\v{r}}{\'\i}},
        title = "{Photometric stellar masses for galaxies in DESI Legacy Imaging Surveys}",
      journal = {\aap},
         year = 2025,
        month = dec,
       volume = {704},
          eid = {A232},
        pages = {A232},
          doi = {10.1051/0004-6361/202453448},
archivePrefix = {arXiv},
       eprint = {2510.02257},
 primaryClass = {astro-ph.GA},
       adsurl = {https://ui.adsabs.harvard.edu/abs/2025A&A...704A.232E}
}

@ARTICLE{Kelvin2014,
       author = {{Kelvin}, Lee S. and {Driver}, Simon P. and {Robotham}, Aaron S.~G. and {Graham}, Alister W. and {Phillipps}, Steven and {Agius}, Nicola K. and {Alpaslan}, Mehmet and {Baldry}, Ivan and {Bamford}, Steven P. and {Bland-Hawthorn}, Joss and {Brough}, Sarah and {Brown}, Michael J.~I. and {Colless}, Matthew and {Conselice}, Christopher J. and {Hopkins}, Andrew M. and {Liske}, Jochen and {Loveday}, Jon and {Norberg}, Peder and {Pimbblet}, Kevin A. and {Popescu}, Cristina C. and {Prescott}, Matthew and {Taylor}, Edward N. and {Tuffs}, Richard J.},
        title = "{Galaxy And Mass Assembly (GAMA): ugrizYJHK S{\'e}rsic luminosity functions and the cosmic spectral energy distribution by Hubble type}",
      journal = {\mnras},
         year = 2014,
        month = apr,
       volume = {439},
       number = {2},
        pages = {1245-1269},
          doi = {10.1093/mnras/stt2391},
archivePrefix = {arXiv},
       eprint = {1401.1817},
 primaryClass = {astro-ph.CO},
       adsurl = {https://ui.adsabs.harvard.edu/abs/2014MNRAS.439.1245K}
}

@ARTICLE{Abdurrouf2022,
       author = {{Abdurro'uf} and {Accetta}, Katherine and {Aerts}, Conny and {Silva Aguirre}, V{\'\i}ctor and {Ahumada}, Romina and {Ajgaonkar}, Nikhil and {Filiz Ak}, N. and {Alam}, Shadab and {Allende Prieto}, Carlos and {Almeida}, Andr{\'e}s and {Anders}, Friedrich and {Anderson}, Scott F. and {Andrews}, Brett H. and {Anguiano}, Borja and {Aquino-Ort{\'\i}z}, Erik and {Arag{\'o}n-Salamanca}, Alfonso and {Argudo-Fern{\'a}ndez}, Maria and {Ata}, Metin and {Aubert}, Marie and {Avila-Reese}, Vladimir and {Badenes}, Carles and {Barb{\'a}}, Rodolfo H. and {Barger}, Kat and {Barrera-Ballesteros}, Jorge K. and {Beaton}, Rachael L. and {Beers}, Timothy C. and {Belfiore}, Francesco and {Bender}, Chad F. and {Bernardi}, Mariangela and {Bershady}, Matthew A. and {Beutler}, Florian and {Bidin}, Christian Moni and {Bird}, Jonathan C. and {Bizyaev}, Dmitry and {Blanc}, Guillermo A. and {Blanton}, Michael R. and {Boardman}, Nicholas Fraser and {Bolton}, Adam S. and {Boquien}, M{\'e}d{\'e}ric and {Borissova}, Jura and {Bovy}, Jo and {Brandt}, W.~N. and {Brown}, Jordan and {Brownstein}, Joel R. and {Brusa}, Marcella and {Buchner}, Johannes and {Bundy}, Kevin and {Burchett}, Joseph N. and {Bureau}, Martin and {Burgasser}, Adam and {Cabang}, Tuesday K. and {Campbell}, Stephanie and {Cappellari}, Michele and {Carlberg}, Joleen K. and {Wanderley}, F{\'a}bio Carneiro and {Carrera}, Ricardo and {Cash}, Jennifer and {Chen}, Yan-Ping and {Chen}, Wei-Huai and {Cherinka}, Brian and {Chiappini}, Cristina and {Choi}, Peter Doohyun and {Chojnowski}, S. Drew and {Chung}, Haeun and {Clerc}, Nicolas and {Cohen}, Roger E. and {Comerford}, Julia M. and {Comparat}, Johan and {da Costa}, Luiz and {Covey}, Kevin and {Crane}, Jeffrey D. and {Cruz-Gonzalez}, Irene and {Culhane}, Connor and {Cunha}, Katia and {Dai}, Y. Sophia and {Damke}, Guillermo and {Darling}, Jeremy and {Davidson}, Jr., James W. and {Davies}, Roger and {Dawson}, Kyle and {De Lee}, Nathan and {Diamond-Stanic}, Aleksandar M. and {Cano-D{\'\i}az}, Mariana and {S{\'a}nchez}, Helena Dom{\'\i}nguez and {Donor}, John and {Duckworth}, Chris and {Dwelly}, Tom and {Eisenstein}, Daniel J. and {Elsworth}, Yvonne P. and {Emsellem}, Eric and {Eracleous}, Mike and {Escoffier}, Stephanie and {Fan}, Xiaohui and {Farr}, Emily and {Feng}, Shuai and {Fern{\'a}ndez-Trincado}, Jos{\'e} G. and {Feuillet}, Diane and {Filipp}, Andreas and {Fillingham}, Sean P. and {Frinchaboy}, Peter M. and {Fromenteau}, Sebastien and {Galbany}, Llu{\'\i}s and {Garc{\'\i}a}, Rafael A. and {Garc{\'\i}a-Hern{\'a}ndez}, D.~A. and {Ge}, Junqiang and {Geisler}, Doug and {Gelfand}, Joseph and {G{\'e}ron}, Tobias and {Gibson}, Benjamin J. and {Goddy}, Julian and {Godoy-Rivera}, Diego and {Grabowski}, Kathleen and {Green}, Paul J. and {Greener}, Michael and {Grier}, Catherine J. and {Griffith}, Emily and {Guo}, Hong and {Guy}, Julien and {Hadjara}, Massinissa and {Harding}, Paul and {Hasselquist}, Sten and {Hayes}, Christian R. and {Hearty}, Fred and {Hern{\'a}ndez}, Jes{\'u}s and {Hill}, Lewis and {Hogg}, David W. and {Holtzman}, Jon A. and {Horta}, Danny and {Hsieh}, Bau-Ching and {Hsu}, Chin-Hao and {Hsu}, Yun-Hsin and {Huber}, Daniel and {Huertas-Company}, Marc and {Hutchinson}, Brian and {Hwang}, Ho Seong and {Ibarra-Medel}, H{\'e}ctor J. and {Chitham}, Jacob Ider and {Ilha}, Gabriele S. and {Imig}, Julie and {Jaekle}, Will and {Jayasinghe}, Tharindu and {Ji}, Xihan and {Johnson}, Jennifer A. and {Jones}, Amy and {J{\"o}nsson}, Henrik and {Katkov}, Ivan and {Khalatyan}, Dr., Arman and {Kinemuchi}, Karen and {Kisku}, Shobhit and {Knapen}, Johan H. and {Kneib}, Jean-Paul and {Kollmeier}, Juna A. and {Kong}, Miranda and {Kounkel}, Marina and {Kreckel}, Kathryn and {Krishnarao}, Dhanesh and {Lacerna}, Ivan and {Lane}, Richard R. and {Langgin}, Rachel and {Lavender}, Ramon and {Law}, David R. and {Lazarz}, Daniel and {Leung}, Henry W. and {Leung}, Ho-Hin and {Lewis}, Hannah M. and {Li}, Cheng and {Li}, Ran and {Lian}, Jianhui and {Liang}, Fu-Heng and {Lin}, Lihwai and {Lin}, Yen-Ting and {Lin}, Sicheng and {Lintott}, Chris and {Long}, Dan and {Longa-Pe{\~n}a}, Pen{\'e}lope and {L{\'o}pez-Cob{\'a}}, Carlos and {Lu}, Shengdong and {Lundgren}, Britt F. and {Luo}, Yuanze and {Mackereth}, J. Ted and {de la Macorra}, Axel and {Mahadevan}, Suvrath and {Majewski}, Steven R. and {Manchado}, Arturo and {Mandeville}, Travis and {Maraston}, Claudia and {Margalef-Bentabol}, Berta and {Masseron}, Thomas and {Masters}, Karen L. and {Mathur}, Savita and {McDermid}, Richard M. and {Mckay}, Myles and {Merloni}, Andrea and {Merrifield}, Michael and {Meszaros}, Szabolcs and {Miglio}, Andrea and {Di Mille}, Francesco and {Minniti}, Dante and {Minsley}, Rebecca and {Monachesi}, Antonela},
        title = "{The Seventeenth Data Release of the Sloan Digital Sky Surveys: Complete Release of MaNGA, MaStar, and APOGEE-2 Data}",
      journal = {\apjs},
         year = 2022,
        month = apr,
       volume = {259},
       number = {2},
          eid = {35},
        pages = {35},
          doi = {10.3847/1538-4365/ac4414},
archivePrefix = {arXiv},
       eprint = {2112.02026},
 primaryClass = {astro-ph.GA},
       adsurl = {https://ui.adsabs.harvard.edu/abs/2022ApJS..259...35A}
}

@ARTICLE{Gadotti2026,
       author = {{Gadotti}, Dimitri A.},
        title = "{Robust galaxy image decompositions with differential evolution optimization and the problem of classical bulges in and beyond the nearby Universe}",
      journal = {\mnras},
         year = 2026,
        month = feb,
       volume = {545},
       number = {4},
          eid = {staf2072},
        pages = {staf2072},
          doi = {10.1093/mnras/staf2072},
archivePrefix = {arXiv},
       eprint = {2511.13823},
 primaryClass = {astro-ph.GA},
       adsurl = {https://ui.adsabs.harvard.edu/abs/2026MNRAS.545f2072G}
}

@INPROCEEDINGS{Silchenko2009,
       author = {{Sil'chenko}, Olga K.},
        title = "{Exponential bulges and antitruncated disks in lenticular galaxies}",
    booktitle = {The Galaxy Disk in Cosmological Context},
         year = 2009,
       editor = {{Andersen}, Johannes and {Nordstr{\"o}ara} and {m}, Birgitta and {Bland-Hawthorn}, Joss},
       series = {IAU Symposium},
       volume = {254},
        month = mar,
        pages = {173-178},
          doi = {10.1017/S1743921308027567},
archivePrefix = {arXiv},
       eprint = {0807.1817},
 primaryClass = {astro-ph},
       adsurl = {https://ui.adsabs.harvard.edu/abs/2009IAUS..254..173S}
}

@ARTICLE{Silchenko2018,
       author = {{Sil'chenko}, Olga K. and {Kniazev}, Alexei Yu. and {Chudakova}, Ekaterina M.},
        title = "{The Structure of Large-scale Stellar Disks in Cluster Lenticular Galaxies}",
      journal = {\aj},
         year = 2018,
        month = sep,
       volume = {156},
       number = {3},
          eid = {118},
        pages = {118},
          doi = {10.3847/1538-3881/aad37b},
archivePrefix = {arXiv},
       eprint = {1809.05202},
 primaryClass = {astro-ph.GA},
       adsurl = {https://ui.adsabs.harvard.edu/abs/2018AJ....156..118S}
}

@ARTICLE{Silchenko2020,
       author = {{Sil'chenko}, Olga K. and {Kniazev}, Alexei Yu. and {Chudakova}, Ekaterina M.},
        title = "{The Structure of Stellar Disks in Isolated Lenticular Galaxies}",
      journal = {\aj},
         year = 2020,
        month = aug,
       volume = {160},
       number = {2},
          eid = {95},
        pages = {95},
          doi = {10.3847/1538-3881/ab9eaf},
archivePrefix = {arXiv},
       eprint = {2007.01129},
 primaryClass = {astro-ph.GA},
       adsurl = {https://ui.adsabs.harvard.edu/abs/2020AJ....160...95S}
}

\end{document}